\documentclass[conference]{IEEEtran}
\IEEEoverridecommandlockouts
\usepackage[table,xcdraw]{xcolor}
\usepackage{cite}
\usepackage{amsmath,amssymb,amsfonts}
\usepackage{algorithmic}
\usepackage{graphicx}
\usepackage{pgf}
\usepackage{mwe}%
\usepackage{booktabs}       
\usepackage{todonotes}
\usepackage{comment}
\usepackage{soul}

\usepackage{verbatim}

\usepackage{url}            

\usepackage{nicefrac}       

\usepackage{textcomp}
\usepackage{svg}
\usepackage{multirow}
\usepackage{tabularx}
\usepackage{graphics}
\usepackage{titlesec}

\usepackage{float}
\usepackage{listings}
\usepackage{tcolorbox}  

\tcbuselibrary{listingsutf8}  

\lstdefinestyle{pythonstyle}{
  frame=single,
  rulecolor=\color{black},
  backgroundcolor=\color{gray!10},
  language=Python,
  basicstyle={\small\ttfamily},
  keywordstyle=\color{magenta},
  commentstyle=\color{blue},
  stringstyle=\color{magenta},
  breaklines=true,
  breakatwhitespace=true,
  tabsize=2
}

\def\BibTeX{{\rm B\kern-.05em{\sc i\kern-.025em b}\kern-.08em
    T\kern-.1667em\lower.7ex\hbox{E}\kern-.125emX}}

\begin{document}

\newtheorem{definition}{Definition}

\newcommand{\pandas}{\texttt{Pandas}}
\newcommand{\cylon}{\textit{Cylon}}
\newcommand{\cflow}{\textit{CylonFlow}}
\newcommand{\pycylon}{\textit{PyCylon}}
\newcommand{\gcylon}{\textit{GCylon}}

\newcommand{\TODO}[1]{\todo[inline]{#1}}
\newcommand\arupNote[1]{\sethlcolor{yellow} \hl{Arup: #1}} %


\title{Combining Serverless and High-Performance Computing Paradigms to support ML Data-Intensive Applications}

\author{
\IEEEauthorblockN{Mills Staylor \IEEEauthorrefmark{2}\IEEEauthorrefmark{4}, 
Arup Kumar Sarker\IEEEauthorrefmark{2}\IEEEauthorrefmark{4},
Gregor von Laszewski\IEEEauthorrefmark{4},
Geoffrey Fox\IEEEauthorrefmark{2}\IEEEauthorrefmark{4},
Yue Cheng\IEEEauthorrefmark{2},
Judy Fox\IEEEauthorrefmark{2},
}

\IEEEauthorblockA{
\IEEEauthorrefmark{2}University of Virginia, Charlottesville, VA 22904, USA
\{qad5gv, djy8hg, mrz7dp, ckw9mp\}@virginia.edu}
\IEEEauthorblockA{\IEEEauthorrefmark{4}Biocomplexity Institute and Initiative, University of Virginia, Charlottesville, VA\\
22911, USA\\
laszewski@gmail.com, vxj6mb@virginia.edu}
}

\maketitle
\begin{abstract}
    Data is found everywhere, from health and human infrastructure to the surge of sensors and the proliferation of internet-connected devices. To meet this challenge, the data engineering field has expanded significantly in recent years in both research and industry. Traditionally, data engineering, Machine Learning, and AI workloads have been run on large clusters within data center environments, requiring substantial investment in hardware and maintenance. With the rise of the public cloud, it is now possible to run large applications across nodes without owning or maintaining hardware. \textbf{Serverless functions such as AWS Lambda provide horizontal scaling and precise billing without the hassle of managing traditional cloud infrastructure. However, when processing large datasets, users often rely on external storage options that are significantly slower than direct communication typical of HPC clusters.  We introduce Cylon, a high-performance distributed data frame solution that has shown promising results for data processing using Python. We describe how we took inspiration from the FMI library and designed a serverless communicator to tackle communication and performance issues associated with serverless functions.
    With our design, we demonstrate that the scaling efficiency of AWS Lambda achieves within 6.5\% of serverful AWS (EC2) at 64 nodes, based on implementing direct communication via NAT Traversal TCP Hole Punching.}
\end{abstract}

\section{Introduction and Motivation}\label{sec:intro}
The data engineering domain has expanded monumentally over the past decade, thanks to the emergence of Machine Learning (ML) and Artificial Intelligence (AI). Data is no longer categorized as Gigabytes (GB) or Megabytes (MB), files, or databases but as terabytes or abstract data stores. It takes a considerable amount of developer time for pre-processing when a better use of time would be designing and implementing deep learning or machine learning models. The increasing number of connected internet devices and social media results in data growing at parabolic or exponential rates.  Improving performance is crucial for building pipelines that support the development of optimized AI/ML training and inference.

The exponential growth in data volume and complexity poses significant challenges across various scientific domains, particularly in genomics, climate modeling, astronomy, and neuroscience. For instance, the 1000 Genomes project and the Cancer Genome Atlas alone utilize nearly five terabytes of data\cite{McKenna2010The}. The TSE search terabase alone consumes a substantial 50.4 TB in size. This can be understood by considering that a single genome sequence alone can generate approximately 200 GB of data\cite{genomefaq:online}. Consequently, by 2025, it will be necessary to store the genome sequences of the entire world, amounting to 40 exabytes of storage.

Decomposing complex problems represented by large datasets can be challenging and demands both scalable and efficient solutions. Machine Learning and Artificial Intelligence have revolutionized data analysis, allowing researchers to extract valuable insights that were previously unattainable. This leads to more robust and reliable results than possible with the previously available tools.

The rise of Big Data and cloud computing, which began in the early 2000s, played a pivotal role in the emergence of modern artificial intelligence and machine learning. Machine learning and deep learning applications rely heavily on large, predetermined, and preprocessed datasets\cite{abeykoon2020data}. The exponential growth of data has posed significant challenges, particularly in terms of traditional storage and the need for efficient data exchange between distributed storage locations. Cloud computing services, such as Amazon Web Services (AWS), offer innovative solutions to these challenges by providing shared resources, including computing, storage, and analytics\cite{islamBigData}.

Function as a Service (FaaS) is a rapidly evolving paradigm in cloud computing applications. In FaaS, users focus on writing code divided into individual functions. This approach allows users to focus on the application code itself, rather than on deploying and managing the underlying compute and storage infrastructure. Serverless functions also provide elasticity and fine-grained billing capabilities. However, serverless currently lacks efficient communication through message passing for large-scale parallel data processing applications, such as Cylon\cite{copik2023fmi}. To address this limitation, Network Address Translation (NAT) TCP Hole Punching offers a high-performance and cost-effective alternative to other methods, such as object or in-memory storage used on AWS Lambda to circumvent communication challenges\cite{moyer2021punching}. Copik et al. detail performance experiments of the FMI library in the paper "FMI: Fast and Cheap Message Passing for Serverless Functions." Performance improvements for FMI usage are observed to be between 105 and 1024 times better than that of AWS DynamoDB. Additionally, the authors detail experiments demonstrating performance converging comparably to FMI and MPI on EC2  around 8 nodes and note improvements in performance for distributed machine learning FaaS applications 162 times and cost reductions up to 397 times\cite{copik2023fmi}.

Hydrology, earthquake prediction, and astronomical image data processing are additional examples of time series foundational models that could be applied and executed on serverless cloud systems like AWS Lambda. For instance, hydrology time series analysis using the CAMELS and Caravan global datasets for rainfall and runoff has been demonstrated with deep learning approaches\cite{he2024science,osti_10634837}.  Similarly, earthquake time series forecasting using MultiFoundationQuake and GNNCoder for the next 14 days using Southern California earthquake data from 1986 to 2024 has been explored\cite{jafari2024time}.  There is existing work to apply and integrate this research in the form of inference and training in Cylon, which could be integrated with Serverless via our proposed communicator integration.  Lastly, Cloud-base Astronomy Inference (CAI), a serverless application of astronomical image data processing, has been demonstrated\cite{Staylor2024cosmic}. This application utilizes the FMI library for collectives and aligns with the proposed AWS Lambda architecture described later. The authors plan to integrate Cylon into CAI in the future and leverage high-performance serverless communication for Machine Learning inference at scale.

There are documented serverless applications for Bioinformatics analysis, leveraging AWS Lambda functions' embarrassingly high parallelism and scalability. Grzesik et al. describe a serverless application that mines single-nucleotide polymorphism (SNP) data. The application pushes data to an Amazon S3 bucket, which is later processed asynchronously \cite{grzesik2022serverless}.  Similarly, Robert Aboukhalil describes serverless executions using \textit{Cloudflare}, a serverless execution service that connects to providers like AWS to simulate DNA sequencing data via streaming to AWS S3 \cite{ServerlessGenomics}.  Hung et al. describe a serverless approach for RNA Sequencing Data analysis. Like other AWS Lambda applications, an S3 bucket is used as an intermediate layer for data transfer. The authors acknowledge design challenges related to the impact of data transfers from the AWS Lambda function to the AWS S3 bucket \cite{hung2020accessible}.  This is the gap our research aims to fill. This work presents a subset of research exploring parallelism and vectorization innovatively across Serverless AWS and HPCs using Apache Arrow. We detail our serverless design that utilizes NAT Traversal TCP Hole Punching as a novel approach to enable Serverless AWS Lambda functions and AWS Fargate for parallel execution on AWS.

In this paper, we will explore our high-performance serverless, a runtime engine designed to optimize the execution of deep learning applications in high-performance computing (HPC) and cloud environments. We introduce a serverless communicator design that leverages NAT Hole Punching to provide a high-performance alternative to intermediate storage solutions like AWS S3.  This enables function-to-function bidirectional communication, a capability not natively supported by AWS Lambda.  This is also an ideal choice for AWS Fargate, serverless Docker on AWS ECS, based on the difficulties in implementing high-performance communication using libraries such as UCX, MPI, and Gloo. We introduce and incorporate Cylon to achieve an existing goal of parallel computing, making it achievable without user intervention. This is accomplished by utilizing libraries of previously parallelized operators (such as join and other Pandas operators, as well as deep learning).  We will explain the advantages of executing workloads within Docker containers on AWS using ECS tasks and discuss how this approach can be extended to HPC environments utilizing Singularity or Apptainer so that it is possible to execute derivative experiments.

Next, we will address the challenges of large-scale parallel execution on AWS by integrating direct communication support via NAT Traversal TCP Hole Punching in Cylon for pre-processing and inference in serverless environments. By demonstrating the effectiveness of our approach, we will highlight its applicability in scientific fields like bioinformatics, hydrology, earthquake prediction, astronomy, and genome research, where it can address high-performance communication issues observed in current state-of-the-art approaches, such as distributed storage latency and data parallelism. Furthermore, we will explore the cost savings associated with FaaS execution to showcase the relevance and innovation of this approach compared to serverful execution in traditional HPC environments.  Our experimental results demonstrate that our serverless communicator design achieves scaling efficiency within 6.5\% of EC2 at 64 nodes for strong scaling experiments, significantly outperforming traditional object storage approaches such as AWS S3 used in current state-of-the-art serverless bioinformatics applications.

\subsection{Contributions}
This work makes the following research contributions:
\newline

\textbf{(i) BSP Execution Model for Serverless Computing:} We demonstrate that Bulk Synchronous Parallel (BSP) workloads, traditionally requiring dedicated high-performance computing (HPC) infrastructure with low-latency interconnects, can effectively execute in serverless environments. By enabling direct peer-to-peer communication between Lambda functions through NAT traversal, we demonstrate that parallel computations achieve scaling efficiency within 6.5\% of that observed on provisioned cloud virtual machines (VMs). This challenges the assumption that serverless computing is limited to embarrassingly parallel or loosely coupled workloads.

\textbf{(ii) Unified Framework for ML Data Pipelines Across Computing Paradigms:} We introduce Cylon, a portable and distributed dataframe library that allows the same data-intensive operations to run seamlessly across various environments, including serverless (AWS Lambda), cloud virtual machines (EC2), and high-performance computing (Rivanna). This capability supports both machine learning training and inference pipelines. 

Scientific fields with significant data processing requirements, such as genomics, hydrology, and astronomy, can benefit from Cylon's distributed operations in serverless deployments. This enables elastic scaling for fluctuating workloads without the need for dedicated infrastructure.

\textbf{(iii) Quantitative Cost-Performance Analysis for Serverless HPC:} We present an empirical cost model for data-intensive serverless workloads. Our analysis shows that serverless can be cost-competitive for bursty workloads: a 32-worker distributed join costs approximately \$0.03 on Lambda with pay-per-millisecond billing, compared with provisioned EC2 instances where idle time dominates cost for intermittent workloads.

\textbf{(iv) Comparison of the communication substrate for Serverless Collectives:} We evaluate three communication approaches for MPI-style collectives in serverless environments: direct TCP through NAT hole-punching, Redis, and AWS S3 message passing. Our results show that direct communication achieves 10-100$\times$ lower latency than storage-mediated alternatives, establishing NAT traversal as a viable approach for latency-sensitive distributed computing in serverless environments.

\section{Related Works} \label{sec:relatedWorks}

One of the fundamental concepts in data science is the dataframe. In the Python ecosystem, Pandas\cite{mckinney2011pandas} was created as a Python derivative of the R \textit{data.frame} class. However, one significant limitation of Pandas is that it can only execute on a single core. In contrast, frameworks like Apache Spark, a Java Virtual Machine (JVM) framework, offer similar capabilities and improved performance and support the dataframe abstraction. One notable drawback of Spark is that it operates as a framework rather than a standalone Python library. While there is support for Python through PySpark, its usage necessitates the configuration, setup, and deployment of a Spark cluster. The JVM-to-Python translations introduced by Spark add a substantial performance bottleneck compared to the C++-to-Python translation implemented by Cylon and other high-performance Python libraries, which are more lightweight\cite{abeykoon2020data}.

Similar to the Cylon library, the Twister2 toolkit developed by Kamburugamuwe et al. is designed using the Bulk Synchronous Parallel (BSP) architecture, based on the observed advantages of scalability and performance. It also incorporates a dataframe API implemented in Java\cite{perera2023depth}. While MPI implementations encompass scalable and efficient communication routines, their process launching mechanisms work well with mainstream HPC systems but are incompatible with some environments that adopt their own resource management systems \cite{shan2022hybrid}. A crucial abstraction introduced by Twister2 is TSets, a concept similar to dataframes or, equivalently, RDDs in Apache Spark or datasets in Apache Flink. Twister2 is considered a foundational step towards high-performance data engineering research and serves as a direct precursor to Cylon\cite{pererathesis}.

The Dask, Modin, and Ray Python distributed dataframe packages are built on the Pandas API and share architectural goals similar to those of the Cylon architecture. A related project, Dask-Cudf, provides distributed dataframe capabilities on Nvidia GPUs, leveraging both Dask and Pandas to facilitate integration.  Perera et al. describe the potential to "supercharge" Dask/Ray performance by integrating Cylon using actors to create a highly efficient HP-DDF runtime\cite{perera2023supercharging}.

To address the need for a uniform distributed architecture, Jeff Dean describes Google’s Pathway architecture, which aims to address the future of AI/ML systems as researchers migrate from Single Program Multiple Data (SPMD) to Multiple Program Multiple Data (MPMD). However, this system is closed-source.  In response, the Cylon Radical Pilot has been designed to support the execution of heterogenous workloads on HPCs. \cite{sarker2024radical}

Rapids is an open-source suite of GPU-accelerated data science and AI libraries developed by Nvidia. The Rapids cuDF dataframe library improves the performance of Pandas operations on GPUs while offering a 100\% compatible API with Pandas, a shared goal with the Cylon project\cite{RAPIDScuDF}.

Another related project, Wukong, offers a similar capability, including support for AI/ML inference, but was built on serverless FaaS where AWS manages provisioning, scaling, and other undifferentiated tasks. This DAG execution framework has demonstrated near-ideal scalability by executing jobs approximately 68 times faster while achieving nearly 92 percent cost savings compared to NumPyWren. Wukong utilizes Redis as a key-value store for intermediate and final storage, addressing the data parallelism and communication performance issues described in related scientific works\cite{carver2020wukong}.

A significant challenge in serverless execution, particularly for parallel data processing applications, is data transfer between running functions. Moyer proposes a solution that leverages Nat Traversal via TCP Hole Punching through an external rendezvous server to enable direct TCP connections between a pair of functions. For an experiment involving over 100 functions, this approach resulted in a performance improvement of 4.7 times compared to using object storage. \cite{moyer2021punching}

A notable difference between Moyer’s work and the FMI library lies in Moyer’s implementation, which utilizes web sockets for communication. In contrast, FMI is designed as a C++ library making it highly applicable to executing parallel processing tasks on serverless architectures such as Cylon.

\section{Design and Implementation}
\label{sec:design}

This section outlines the overarching design of Cylon, the foundation of our High-Performance Serverless Architecture. The design is shaped by insights gained from developing the Twister2 toolset. Specifically, Cylon aims to achieve the following goals: 1) High performance and scalability, 2) An extensible architecture, 3) User-friendly APIs, and 4) Integration across multiple platforms. Following this overview, we will shift to an analysis of our AWS stack's semantics, which builds on this groundwork. We explore the challenges and architecture of serverless data engineering at scale, identify key pillars and characteristics of cloud architectures, and describe our high-performance communication layer, which builds upon FMI and extends this library to enable the combination of serverless and high-performance computing on serverless platforms \cite{pererathesis}.

\subsection{Cylon Library Architecture/Design}

Cylon was developed to address shortcomings, interact directly with high-performance kernels, and perform better than libraries such as Pandas.  Cylon represents an architecture where performance-critical operations are moved to a highly optimized library. Moreover, the architecture can take advantage of the performance benefits associated with in-memory data and distributed operations and data across processes, an essential requirement for processing large data engineering workloads at scale. Such benefits are realized, for example, in the conversion from tabular or table format to tensor format required for Machine Learning/Deep Learning or via relational algebraic expressions such as joins, select, project, etc. A key aspect of the Cylon architecture is the intersection of data engineering and AI/ML, allowing it to interact seamlessly with frameworks such as Pytorch\cite{pytorch2019} and TensorFlow\cite{tensorflow2015}.

\begin{figure}[H]
    \begin{center}
    \includegraphics[width=\linewidth]{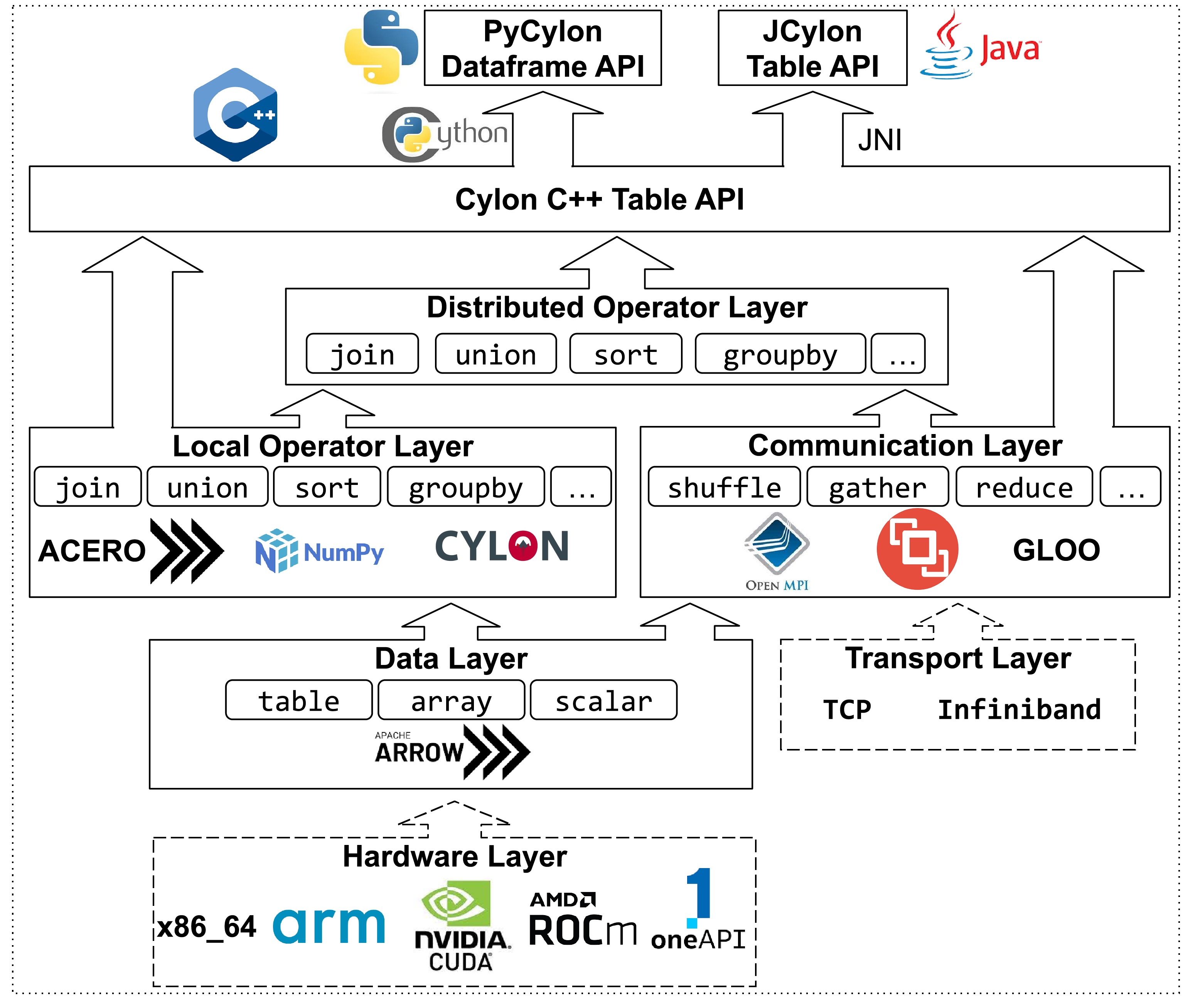}
    \end{center}
    \caption{Serverless Architecture \cite{sarker2024radical} }
     \label{fig:cylonarch}
\end{figure}

Figure \ref{fig:cylonarch} provides a high-level overview of our architecture. Solid boxes represent features and capabilities implemented within the Cylon runtime. The Apache Arrow runtime manages tables, arrays, and scalars, forming the table layer. Our choice of Apache Arrow stems from several factors:
\begin{enumerate}
    \item Many high-performance computing (HPC) libraries utilize C++ (e.g., MPI).
    \item Python-based libraries are widely used in data science and machine learning/artificial intelligence.
    \item We need to transfer data between these environments while maintaining network integration efficiently.
\end{enumerate}

Apache Arrow offers several advantages that align with our requirements:
\begin{enumerate}
    \item Interoperability is provided.
    \item Performance is achieved through in-memory processing and zero-copy sharing.
    \item Communication is facilitated through data transfer and streaming.
    \item Projects like Spark, BigQuery, and Data Fusion widely adopt Apache Arrow.
\end{enumerate}

Cylon supports Parquet integration, a file format that optimizes data storage by using a columnar format instead of a row format. Arrow facilitates loading Parquet files into Arrow for rapid, in-memory processing and then saving the results back into Parquet for long-term storage.

The communication layer represents the interface between Cylon and supported communication libraries. Currently, Cylon supports OpenMPI, UCX \ \cite{shan2022hybrid}, Gloo, and our contribution, FMI. Both distributed and local operators are supported. Cython bindings facilitate Python access to C++ data structures, functions, and classes.

\begin{figure}[H]
    \begin{center}
    \includegraphics[width=\linewidth]{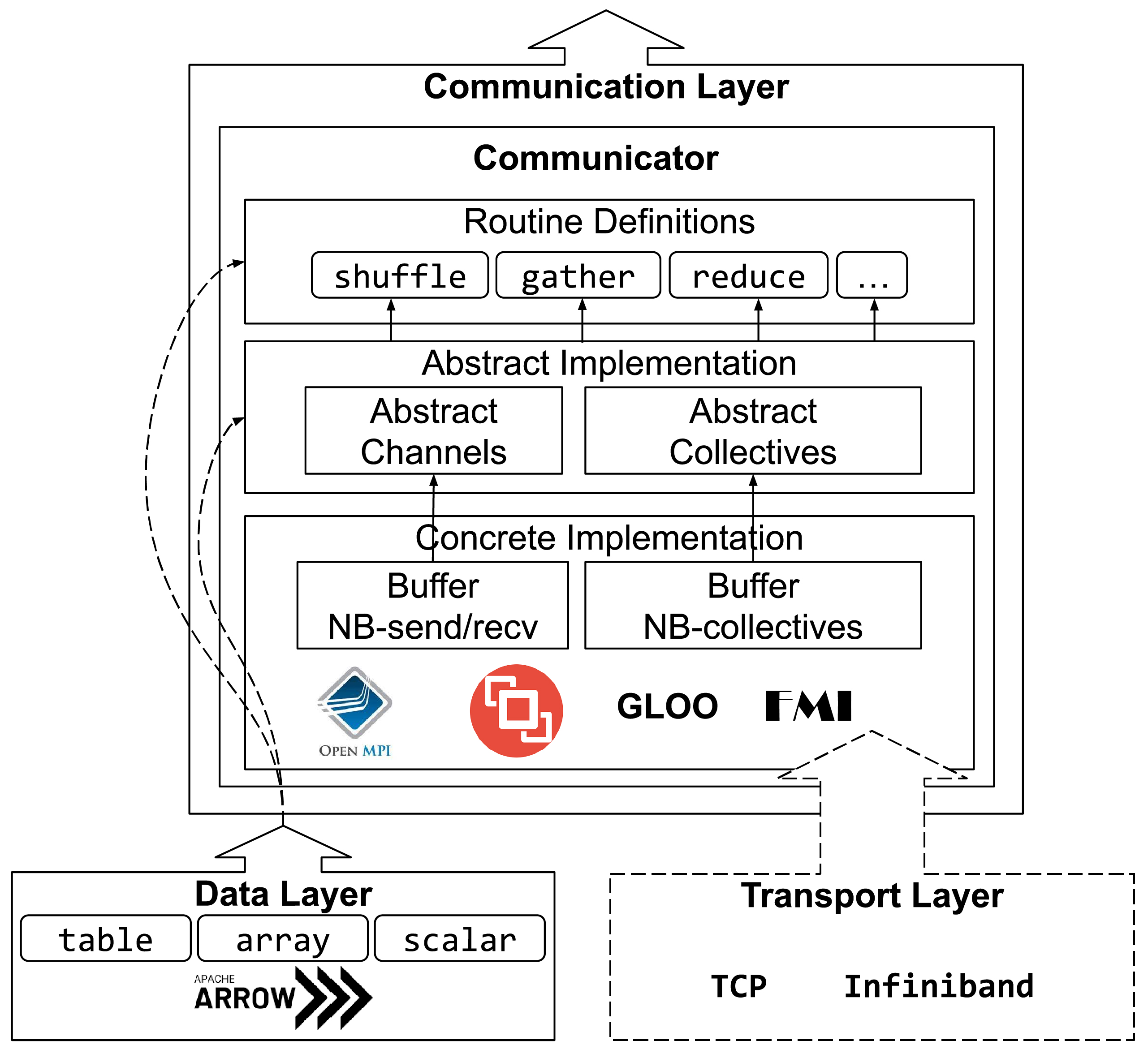}
    \end{center}
    \caption{Communication Model! }
    \label{fig:cyloncomm}
\end{figure}

Figure \ref{fig:cyloncomm} illustrates the communication interface used by our high-performance Python/C++ runtime. Two key features of this model are its modular architecture and extensibility, as discussed in Perera's thesis, "Towards Scalable High Performance Data Engneering Systems."  To simplify the complexities of distributed programming, frameworks like TCP and Infiniband facilitate plug-and-play configuration, supporting high-performance communication libraries such as OpenMPI, UCX, and Gloo. The architecture implements in-memory tables as supported by the Apache Arrow Columnar format. This library also offers serialization-free copying across various runtimes, including Pandas and Numpy, as mentioned in "Towards Scalable High Performance Data Engineering Systems" \cite{pererathesis}. We anticipate that this library will continue to evolve and will profoundly impact the development of high-performance MLSys frameworks and ultimately contribute to and help shape the integration of ML/AI pipelines. 

\subsection{Pillars of Cloud Computing and Well-Architected Cloud Systems}
Cloud computing is widely regarded as the cornerstone of the fourth industrial revolution. As companies and organizations increasingly rely on Big Data and Artificial Intelligence, cloud providers like AWS are pivotal in facilitating integration and innovation to support these endeavors \cite{hashemipour2020amazon}. AWS offers four fundamental pillars for successful cloud architectures. First, cloud architectures eliminate the uncertainty surrounding capacity requirements. Second, they provide virtually unlimited usage, enabling architectures to scale up and down autonomously. Third, cloud architectures facilitate testing systems at production scale. In other words, production-like environments can be created on demand, simulating production environment conditions. Cloud architectures also lower the risk associated with architectural change by removing the test serialization problem with on-premise infrastructure.  This refers to the situation that can occur with software testing on-premise.  In this case, tests are often executed in parallel due to limitations with shared resources or hardware constraints.  This can slow testing and create other bottlenecks that complicate testing and evaluation.  To address this problem, cloud systems support automation as an enabler of architectural experimentation. This allows for the creation and replication of systems at low costs and enables tracking and auditing of impact with low effort.  Lastly, cloud architectures allow for evolutionary architectures.  This includes automating and testing on-demand to lower the impact of design change.  With evolutionary architectures, systems can evolve so that it is possible to take advantage of innovation(s) as standard practice\cite{aws_well_architected_framework}.

Well-architected cloud systems are built on four key pillars: security, reliability, performance efficiency, and cost optimization\cite{aws_well_architected_framework}.

Security protects information systems and assets while delivering value through risk assessment and mitigation strategies. It involves applying security at all layers, enabling traceability, automating responses to security events, and treating automation as a best practice\cite{aws_well_architected_framework}.

Reliability refers to a system’s ability to recover from infrastructure or service failures, acquire resources dynamically to meet demand, and mitigate disruptions such as misconfiguration or transient network issues. Design principles of reliability include testing recovery procedures, automating recovery from failures, using horizontal scalability to increase aggregate system availability, and eliminating guesswork about capacity\cite{aws_well_architected_framework}.

Performance Efficiency denotes the ability to use computing resources efficiently to meet system requirements and maintain that efficiency as demand changes and technology evolves. Key principles related to performance efficiency include democratizing advanced technologies, going global in minutes, and experimenting often\cite{aws_well_architected_framework}.

Cost Optimization involves avoiding unnecessary costs and suboptimal resource use. Effective cost optimization includes using managed services to reduce the overall cost of ownership, trading capital expenses for operating expenses, taking benefits from economies of scale, and reducing spending on data center operations\cite{aws_well_architected_framework}.

This research aligns with the best practices outlined in the discussed pillars of successful cloud architectures and the key characteristics of well-architected cloud architectures. 

\subsection{Serverless, High-Performance Architecture Using Cylon}
As part of this work, we have designed a high-performance serverless architecture within Cylon based on a reference FMI interface to support the direct communication necessary for Bulk Synchronous Parallel architectures. As part of this work, We will discuss a reference serverful architecture for running data-parallel applications on HPCs and traditional cloud services such as EC2 using Cylon, and outline our design that introduces a high-performance serverless communicator that can be used with serverless infrastructure such as AWS Fargate and AWS Lambda.

\subsection{Serverful High Performance Communication using UCC/UCX and Cylon}

Data parallelism in our design is built around Bulk Synchronous Parallel (BSP), a model that uses Single Program Multiple Data (SPMD) across compute nodes.  The Message Passing Interface (MPI) is implemented by frameworks such as OpenMPI, MPICH, MSMPI, IBM Spectrum MPI, etc.

For most serverful use cases, parallel data processing with OpenMPI is standard, based on the OpenMPI library’s availability on HPCs like Rivanna and Summit. However, MPI isn’t compatible with all frameworks, such as Dask and Ray. These distributed libraries are better suited for cloud technologies like Kubernetes and Amazon Elastic Container Service (ECS). OpenMPI includes process bootstrapping capabilities that facilitate the initialization of communication primitives and configuration. Cloud environments like AWS don't require this process management capability; therefore, communication libraries such as UCX/UCC are ideal. UCX is an open-source, production-grade communication framework for data-centric and high-performance applications\cite{UCXRepo}.   We use UCX for point-to-point communication.  UCC is a library and API for collective communication operations that is flexible, complete and feature rich for current and emerging programming models and runtimes\cite{UCCRepo}.  We use UCC for collective operations.   A key benefit of UCX is the library is designed to be decoupled from network hardware.  This provides increased portability while ensuring high performance and scalability.  Another key advantage of UCX is that it allows access to GPU memory and bidirectional GPU communication.  This library also supports Remote Direct Memory Access (RDMA) over Infiniband and Converged Ethernet (ROcE).  Zero-copy GPU memory copy over RDMA is also supported, leading to exceptional performance.  OpenSHEM is also included to support parallel programming\cite{shan2022hybrid}.

For Cylon and as the predecessor of our proposed serverless work, we implemented a configuration mechanism that is modularized in such a way as to enable the implementation of new sources as they become available. We used Redis as a key-value store for our experiments to facilitate the bootstrapping process and provide barrier conditions.
Another interesting detail about this implementation concerns resource usage, specifically with Unified Communication Protocols (UCP) or endpoints for a network connection.  UCC also uses endpoints for collective operations. Unfortunately, reusing endpoints is impossible; therefore, we implemented separate endpoints for both UCC and UCP. This feature is not included with UCC/UCX and is necessary for parallel pre-processing tasks and BSP style execution using Cylon. The integration of UCX as a Cylon communicator involves the following: 1) Serializer, 2) Communication Operators, 3) Channels and AllToAll, and 4) Integration with dataframe operators\cite{shan2022hybrid}.  

\begin{figure}[ht]
    \begin{center}
    \includegraphics[width=\linewidth]{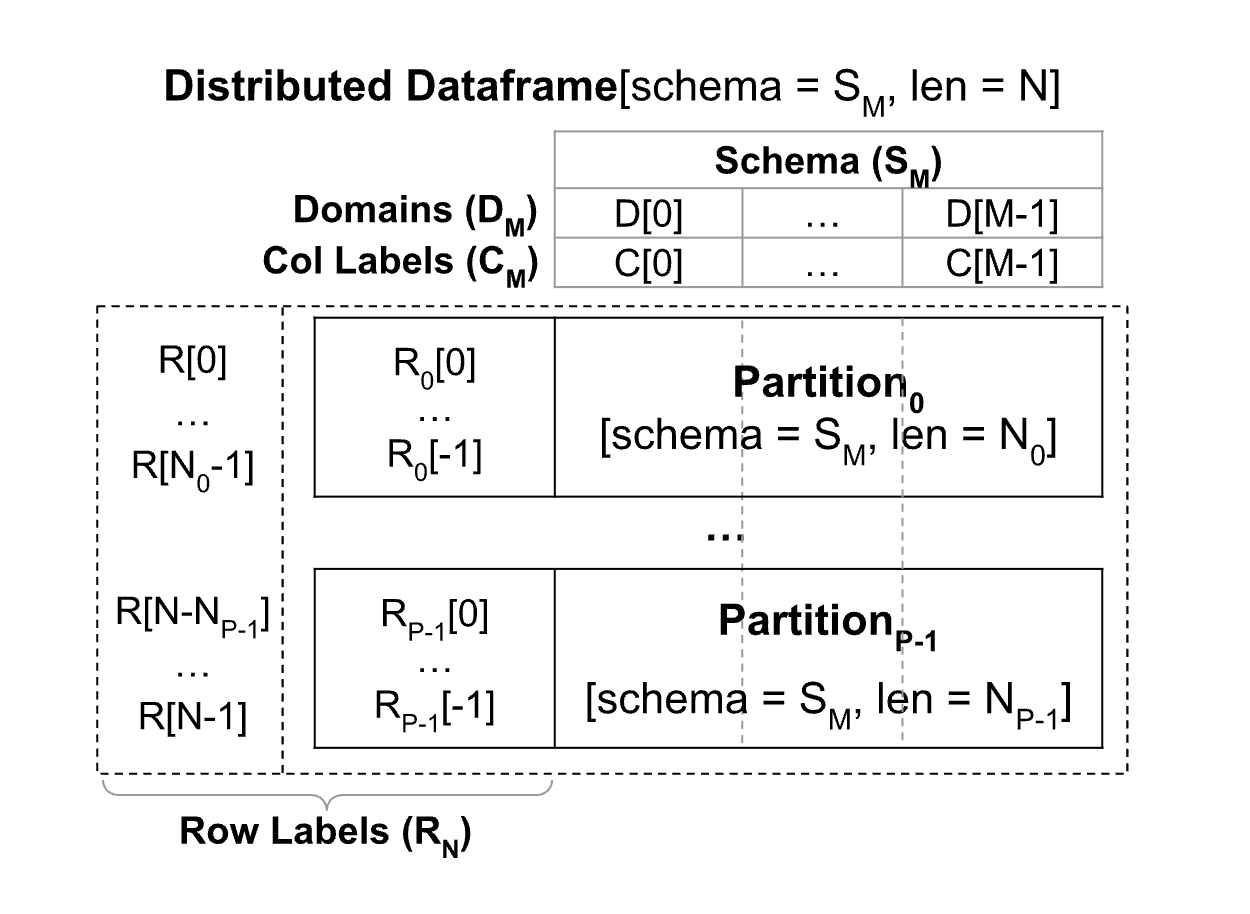}
    \end{center}
    \vspace{-.5cm}
    \caption{Distributed Memory Dataframe (DDMF)\cite{pererathesis}}
    \label{fig:distdataframe}
\end{figure}

Within Cylon, the UCX communicator serializes tables, columns, and scalars into buffers before being passed to the network communication layer.  As was mentioned earlier, Cylon implements a Distributed Memory Dataframe (DDMF) via Apache Arrow. This is represented as a collection of \textit{P} dataframes or partitions of lengths $\{ N_0,...N_{P-1}\}$ and a schema denoted as $S_m$.  The total length of the Distributed Memory Dataframe can be represented as $\sum{N_i}$ and the concatenation of row labels as $R_n=\{R_0, R_1, ... R_{P-1}  \}$.  Figure \ref{fig:distdataframe} illustrates this process.

The Cylon channels API implements the AllToAll operation as point-to-point communication and uses the UCC library for  collective operations such as AllGather.  Distributed operators are implemented generically within Cylon.  The experiments detailed in the next section use the Distributed Join DataFrame operator.  For this case, the process follows: 1) Hash applicable columns into partitioned tables, 2) Use AllToAll to send tables to the intended destination, and 3) Execute a local join on the received tables\cite{shan2022hybrid}.

The initialization or bootstrapping of the Cylon UCX/UCC communicator necessitates the provision of communication metadata during startup. This metadata includes each process's world size, rank, and endpoint address. We utilize the \textit{Redis} key-value store to facilitate this exchange. Our design increments an atomic value to represent the rank. Before incrementing the value, the rank is set. Redis is launched externally to the HPC processes. The only requirement is that the \textit{Redis} host be provided before initializing the HPC runtime. Our experiments employed a combination of AWS ElastiCache, Redis, and ECS tasks, including the dockerized \textit{Redis} runtime. Depending on the specific experiment’s communication requirements, we selected a combination of \textit{Redis} deployments. One implementation limitation is that the stored metadata on \textit{Redis} must be cleared between subsequent experiments. Otherwise, the experiment executes non-deterministically and ultimately fails. In the case of executing N experiments of a given class (e.g., world size of 8 for 9.1 million rows), we need to configure N \textit{Redis} hosts when running experiments in parallel.

\begin{figure}[ht]
    \begin{center}
    \includegraphics[width=\linewidth]{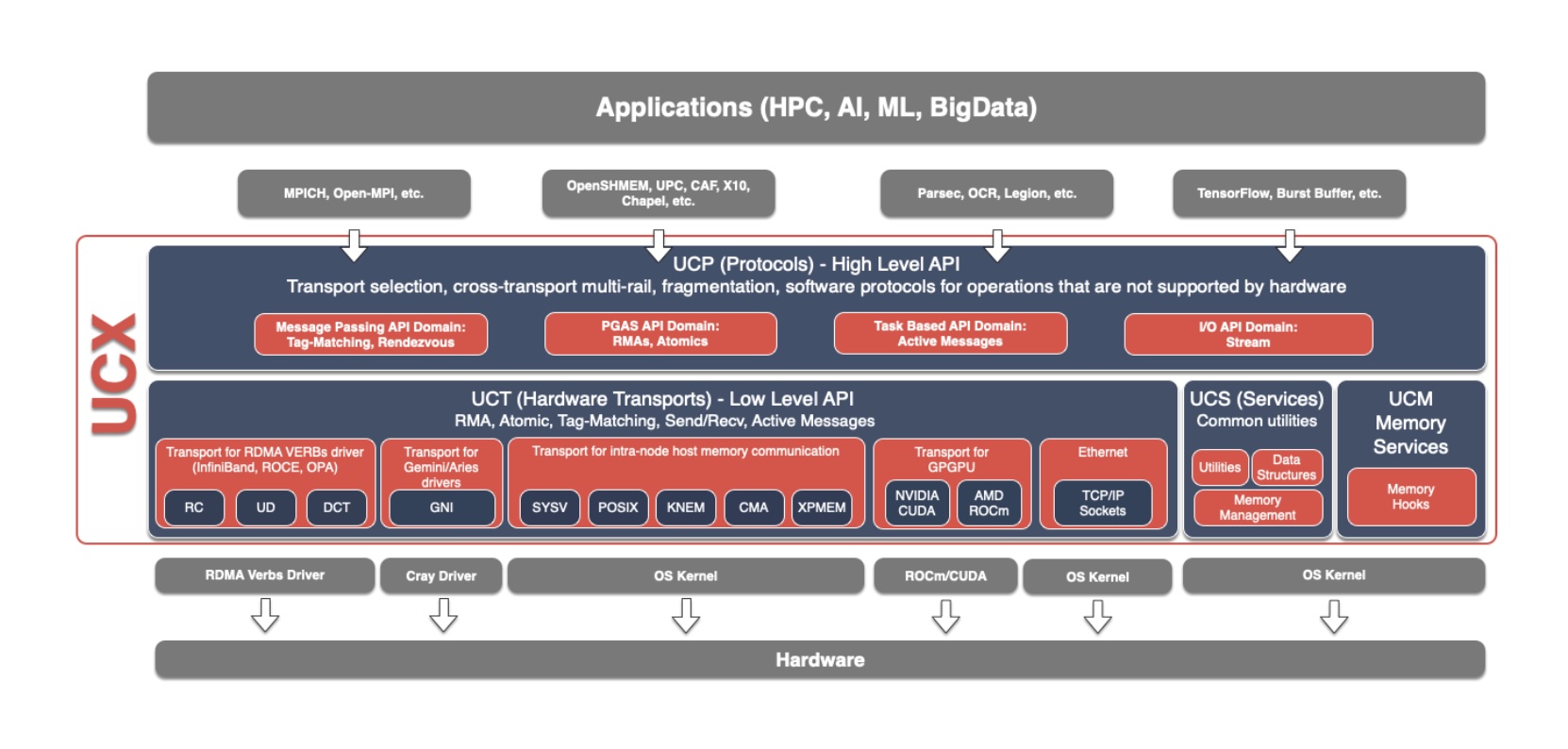}
    \end{center}
    \caption{Unified Communication X (UCX) Library Architecture\cite{shamis2015ucx}}
    \label{fig:ucxarch}
\end{figure}

Our experiment design, which utilized UCX/UCC for our initial experiments, proved highly effective for most AWS and HPC experiments. One aspect of our design involved using Docker. Docker uses OS-level virtualization to deploy software in packages or containers to create a runtime environment that could be seamlessly deployed across both HPC and AWS\cite{dockerwiki}. We pushed our Dockerized environments to Dockerhub and employed Apptainer on Rivanna to construct a Singularity container encapsulating the Cylon runtime. A significant advantage of this approach is that modifications to the environment (or Dockerfile) were not required. Consequently, all our experiments utilized the same runtime, resulting in a consistent and derivative approach across infrastructure.  We believe our architecture and the inclusion of UCX/UCC to be a novel contribution on its own, and based on our research, it is the only case we could find where UCC/UCX was being used for ML/AI pipelines such as Cylon.

\subsection{Serverless High-Performance BSP Architecture}

A crucial aspect of our research involves executing data engineering pre-processing on serverless compute. For our initial exploratory work, we modified the UCX Unified Communication Transport (UCT) lower-level API. This change was necessary due to UCX’s initialization behavior. UCX employs low-level system calls to select the most suitable and optimal communication protocol based on the available resources. However, only TCP is supported for serverless computing. We observed that the UCX device query incorrectly retrieved the wrong IP address from the hardware during initialization. To ensure successful initialization, we had to override the IP address discovered by UCX. This modification enabled us to conduct experiments on AWS Fargate, a serverless Docker running on ECS, a fully managed container orchestration service. Unfortunately, this capability was deprecated due to changes in the underlying serverless architecture implemented by AWS. This result led us to investigate other potential communication options for BSP systems that operate on serverless systems and represents this research.

\begin{figure}[ht]
    \begin{center}
    \includegraphics[width=\linewidth]{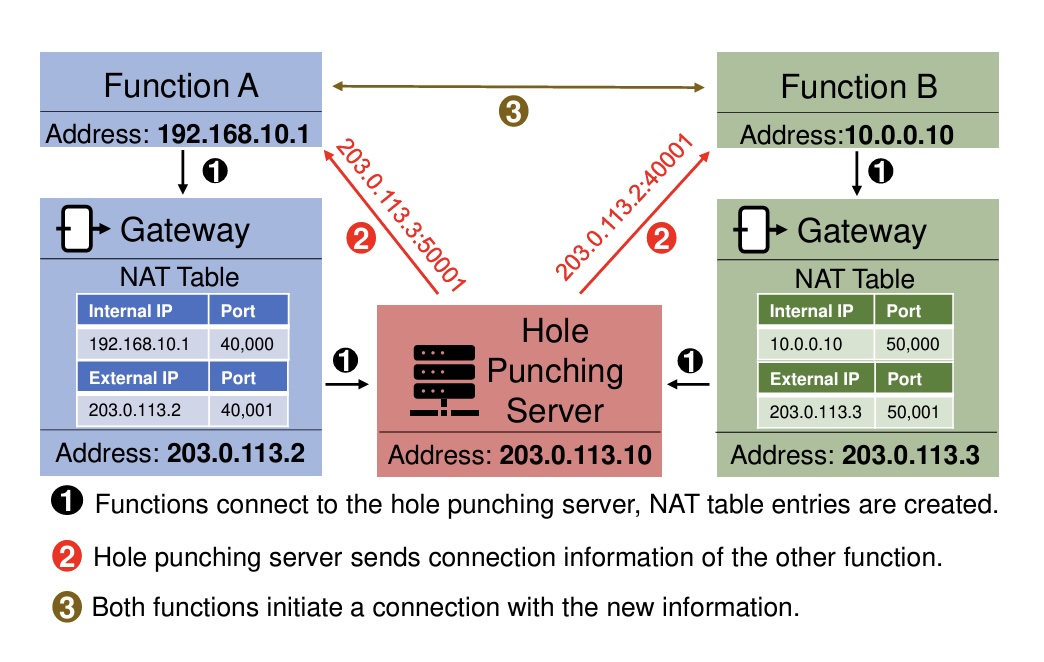}
    \end{center}
    \vspace{-0.5cm}
    \caption{Network Address Translation (NAT) Hole Punching\cite{copik2023fmi}}
    \label{fig:natholepunching}
\end{figure}

A similar issue was encountered with AWS Lambda functions: we could not leverage the IP address override capability mentioned above to execute our HPC runtime on AWS Lambda due to the unavailability of system calls, a similar issue observed on AWS Fargate. One solution presented by Copik et al. was to use NAT traversal or TCP hole punching to enable direct communication between a pair of AWS Lambda functions via the FMI library. Unlike storage-based solutions, direct communication reduces communication time by multiple orders of magnitude.  Figure \ref{fig:natholepunching} illustrates this process.  NAT hole punching requires functions to be behind a NAT gateway.  The gateway hides the addresses of each function by rewriting the internal addresses with an external address.  To support NAT Hole Punching, a publicly accessible server must create entries in the translation table and relay addresses between two functions.

Our initial plan was to design our serverless architecture to support NAT Hole Punching by modifying the existing TCP UCT UCX source code. However, this approach proved unsuccessful due to the UCX TCP architecture and NAT hole-punching address exchange semantics.

A more suitable approach is to design a custom communicator within our communication api, similar to the UCC/UCX communicator described earlier. During our initial work that led to this work, we successfully recreated and validated the FMI library’s functionality as a custom AWS Python Lambda Docker image, and used this as a reference architecture.

To achieve this, we created a Dockerfile that included all the necessary libraries for executing experiments. We also implemented the AWS Lambda interface and developed derivative experiments, such as the distributed join experiments detailed in the next section.

Including this library is an ideal choice, considering the limitations imposed by AWS Lambda. A higher-level abstraction is necessary for direct communication between Lambda functions. Furthermore, there are cost considerations based on the AWS Lambda cost model, where the request rate, execution time, and function size directly impact the overall cost. By supporting direct communication in serverless, we can save costs associated with data storage compared to the current state-of-the-art for serverless applications in domains like bioinformatics. These applications often require parallel, high-performance analysis that would benefit from high-performance communication. Based on results published in the paper, "FMI: Fast and Cheap Message Passing for Serverless Functions," we were able to achieve similar savings as reported in the paper (i.e., 397 times savings compared to the use of mediated storage for Machine Learning applications reported by Copik et al.)\cite{copik2023fmi}.

\begin{figure}[ht]
    \begin{center}
    \includegraphics[width=\linewidth]{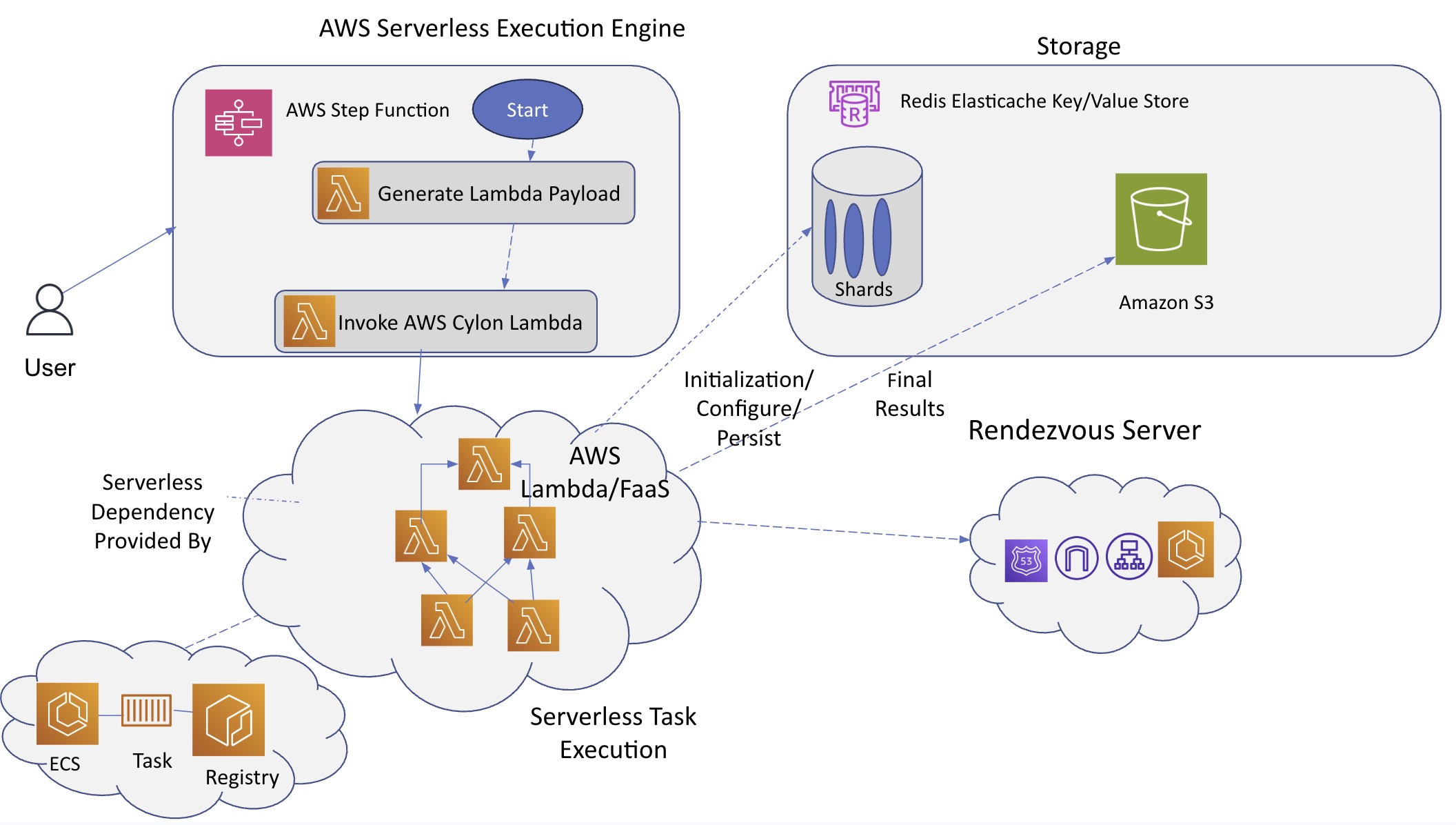}
    \end{center}
    \caption{Serverless Data Engineering Architecture}
    \label{fig:serverlessdataengineering}
\end{figure}

\begin{figure}[ht]
    \begin{center}
    \includegraphics[width=0.80\linewidth, trim={1cm 0.2cm 2cm 0.2cm}, clip]{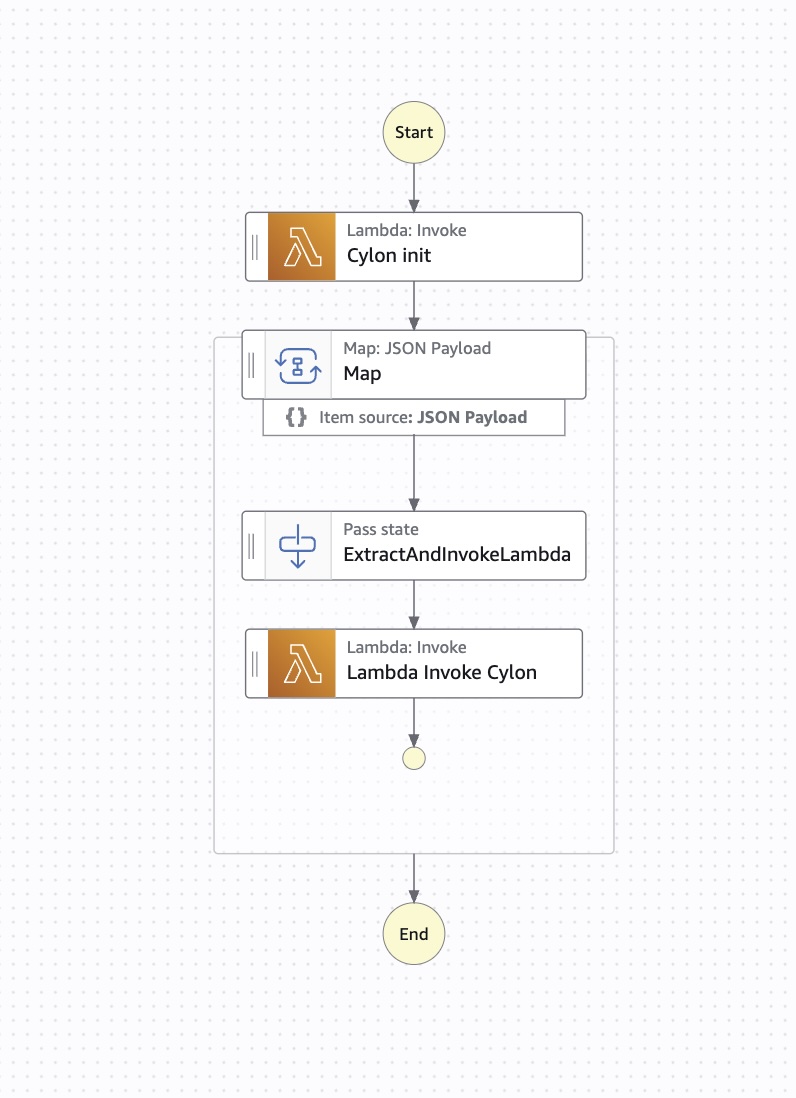}
    \end{center}
    \caption{AWS Step Function for Serverless Data Parallel Task}
    \label{fig:fmistatemachine}
\end{figure}

Figure \ref{fig:serverlessdataengineering} illustrates our architecture. Our design calls for the use of an AWS Step Function to encapsulate the AWS Serverless Execution Engine, as shown in Figure \ref{fig:fmistatemachine}. AWS Step Functions are a Software as a Service (SaaS) feature that facilitates the orchestration of AWS services into serverless workflows. The \textit{Lambda: init} function receives a payload containing various parameters, such as the number of rows, world size, number of iterations, target S3 bucket, script, and output paths. We utilize a specific S3 bucket to store invocation results in a folder using the \textit{Boto3} library. This approach is advantageous over parsing the data output from log files in AWS CloudWatch, a SaaS metrics repository created and utilized by AWS services, as it would be cumbersome. The Lambda invoke state forwards this input JSON payload to a \textit{cylon\_init} AWS Lambda function. This function validates the availability of the rendezvous server. It generates an N-sized array of JSON payloads, which is then sent to the \textit{Map}, \textit{ExtractAndInvokeLambda} states, which is configured to use distributed processing mode to enable highly parallel execution. Next, each generated payload is forwarded to the Lambda: Invoke state from the pass \textit{ExtractAndInvokeLambda} state.  This state passes the input to output, executing the \textit{cylon\_executor} AWS Lambda function. The runtime of the \textit{cylon\_executor} function includes the high-performance MLSys runtime described earlier, which consists of a communicator that supports direct communication over a NAT gateway described earlier. The incoming payload specifies the location of the target script in the S3 bucket. Upon execution of the Lambda function, the contents of the specified S3 bucket are retrieved and persisted in temporary ephemeral storage. Subsequently, Python is used to execute the script used for our experiments. Our design supports both a single script and a folder containing Python scripts and modules.

During execution, the \textit{cylon\_executor} connects with the Rendezvous Server referenced in Figure \ref{fig:serverlessdataengineering}.  Once the second function in the pair connects to the Rendezvous server, both servers receive the destination function's public address.  The Python script continues to run until the collective operations are performed.  Our design also includes using the cloudmesh stopwatch library to facilitate logging and benchmarking the execution time of our experiments\cite{cloudmeshcommon}.
\section{Experiments}

As part of this research, we conducted microbenchmark experiments for the distributed join operation detailed in Table \Ref{tab:exp_table1}. The size of the infrastructure relative to memory and CPU cores was configured based on the memory available to AWS Lambda functions, which has a maximum of 10240 MB. Therefore, we configured a maximum and a close to half-size configuration for each set of experiments to be 10 GB and 6 GB.

\begin{table}[!hbtp]
    \centering
    \small
   
    \caption{Experiments Setup for Comparative to AWS Lambda on UVA Rivanna HPC and Amazon Web Services. WS/SS = weak/strong 
         scaling; M=Million}
        \begin{tabular}{|>{\centering\arraybackslash}p{2.2cm}|>{\centering\arraybackslash}p{2.8cm}|>{\centering\arraybackslash}p{2cm}|} 
        \hline
        \textbf{Experiment Description} & \textbf{Rows} & \textbf{Rows Size} \\
        \hline
        AWS ECS EC2 4 CPU/15 GB Mem WS/SS & 1-64 & [9.1 | 4.5M]  \\
        \hline
        AWS ECS EC2 2 CPU/7.5 GB Mem WS/SS & 1-64 & [9.1 | 4.5M]  \\
        \hline
        Rivanna 10 GB Mem WS/SS & 1-64 & [9.1 | 4.5M]  \\
        \hline
        Rivanna 6 GB Mem WS/SS & 1-64 & [9.1 | 4.5M]  \\
        \hline
        Lambda 10 GB Mem WS/SS & 1-64 & [9.1 | 4.5M]  \\
        \hline
        Lambda 6 GB Mem WS/SS & 1-64 & [9.1 | 4.5M]  \\
        \hline
    \end{tabular}
    \label{tab:exp_table1}
\end{table}
\begin{lstlisting}[
style=pythonstyle, float, 
basicstyle=\footnotesize\ttfamily,
caption=Python Scaling Join Experiment]
import pandas as pd
from numpy.random import default_rng

init_start = time.time()

    if data['env'] == 'fmi':
        communicator = fmi_communicator(data)
        rank = int(data["rank"])
        world_size = int(data["world_size"])
    elif data['env'] == 'fmi-cylon':
        communicator, env = cylon_communicator(data)
        rank = env.rank
        world_size = int(data["world_size"])
    else:
        communicator, env = cylon_communicator(data)
        rank = env.rank
        world_size = env.world_size

    init_end = time.time()

    com_init = (init_end - init_start) * 1000

    u = data['unique']
...
    for i in range(data['it']):

        barrier_start1 = time.time()

        if data['env'] == 'fmi':
            barrier(communicator)
        else:
            barrier(env)
        #print("passed barrier")
        barrier_time1 = (time.time() - barrier_start1) * 1000
        ta['it']}")
        t1 = time.time()

        if data['env'] == 'fmi':
            df3 = pd.concat([df1, df2], axis=1)
            result_array = df3.to_numpy().flatten().tolist()

        else:
            df3 = df1.merge(df2, on=[0], algorithm='sort', env=env)
...
\end{lstlisting} 
\begin{table}
	\centering
	\caption{Execution Time of Weak Scaling from Join Operations Across Infrastructure.}
	\label{tab:weak-exp_table}
	\begin{tabular}{llcr @{\hspace{1\tabcolsep}} lr @{\hspace{1\tabcolsep}} l}
		\toprule
		                   &
		               &
		                     &
		\multicolumn{2}{c}{Execution Time}         \\
		Infrastructure             &
		Scaling               &
		Parallelism                     &
		\multicolumn{2}{c}{time (seconds)}                    \\
		\midrule
		\multirow{5}{*}{EC2 15GB Mem 4 vCPUs} &
		\multirow{5}{*}{Weak} &
		1                       &
		31.57 & $\pm0.20$     \\
		&
		                     &
		2                     &
		40.42 & $\pm1.09$   \\
		&
		                   &
		4                     &
		42.48 & $\pm1.67$    \\
		                    & &
		8                     &
		44.08 & $\pm1.50$      \\
		                    & &
		16                     &
		47.84 & $\pm0.78$      \\
        & &
		32                     &
		49.83 & $\pm0.47$      \\
        & &
		64                     &
		52.70 & $\pm0.34$      \\

        \cmidrule{2-5}
    	\multirow{5}{*}{EC2 7.5GB Mem 2 vCPUs} &
		\multirow{5}{*}{Weak} &
		1                       &
		31.71 & $\pm0.48$     \\
		&
		                     &
		2                     &
		43.63 & $\pm0.43$   \\
		&
		                   &
		4                     &
		46.56 & $\pm0.31$    \\
		                    & &
		8                     &
		49.11 & $\pm0.56$      \\
		                    & &
		16                     &
		51.12 & $\pm0.22$      \\
        & &
		32                     &
		50.97 & $\pm0.30$      \\
& &
		64                     &
		54.98 & $\pm0.49$      \\
		\cmidrule{2-5}
    	\multirow{5}{*}{Lambda 10 GB Mem} &
		\multirow{5}{*}{Weak} &
		1                       &
		30.29 & $\pm1.72$     \\
		&
		                     &
		2                     &
		42.04 & $\pm0.63$   \\
		&
		                   &
		4                     &
		44.93 & $\pm0.16$    \\
		                    & &
		8                     &
		51.13 & $\pm0.51$      \\
		                    & &
		16                     &
		56.52 & $\pm0.88$      \\
        & &
		32                     &
		60.86 & $\pm0.29$      \\
        & &
		64                     &
		64.58 & $\pm2.20$      \\
		
		\cmidrule{2-5}
    	\multirow{5}{*}{Lambda 6 GB Mem} &
		\multirow{5}{*}{Weak} &
		1                       &
		33.31 & $\pm0.34$     \\
		&
		                     &
		2                     &
		44.08 & $\pm0.79$   \\
		&
		                   &
		4                     &
		46.93 & $\pm0.72$    \\
		                    & &
		8                     &
		50.98 & $\pm0.432$      \\
		                    & &
		16                     &
		56.06 & $\pm0.98$      \\
& &
		32                     &
		60.62 & $\pm0.92$      \\
        & &
		64                     &
		64.07 & $\pm2.46$      \\
        \cmidrule{2-5}
    	\multirow{5}{*}{Rivanna 10 GB Mem} &
		\multirow{5}{*}{Weak} &
		1                       &
		18.24 & $\pm0.04$     \\
		&
		                     &
		2                     &
		20.60 & $\pm0.12$   \\
		&
		                   &
		4                     &
		20.78 & $\pm0.04$    \\
		                    & &
		8                     &
		21.40 & $\pm0.37$      \\
		                    & &
		16                     &
		23.05 & $\pm0.50$      \\
                            & &
		32                     &
		24.03 & $\pm0.25$      \\
                            & &
		64                     &
		36.92 & $\pm1.70$      \\
        
        \cmidrule{2-5}
    	\multirow{5}{*}{Rivanna 6 GM Mem} &
		\multirow{5}{*}{Weak} &
		1                       &
		18.27 & $\pm0.10$     \\
		&
		                     &
		2                     &
		20.60 & $\pm0.11$   \\
		&
		                   &
		4                     &
		20.72 & $\pm0.22$    \\
		                    & &
		8                     &
		21.42 & $\pm0.33$      \\
		                    & &
		16                     &
		23.05 & $\pm0.34$      \\
                            & &
		32                     &
		24.89 & $\pm1.26$      \\
                            & &
		64                     &
		36.14 & $\pm0.79$      \\
		\bottomrule
	\end{tabular}
\end{table}
\subsection{Join Operation Scalability}

For these experiments, we ran scaling experiments plotted as speedup on AWS Lambda, EC2 ECS clusters and Rivanna.  The experiment design involved running experiments on derivative-configured infrastructure.  For EC2,  we used Ubuntu 22.04.2 LTS (Intel(R) Xeon(R) CPU E5-2670 v2 @ 2.50GHz) for hosts configured with two virtual CPU and 15 GB of memory (m3xlarge) and for hosts configured with one virtual CPU and 7.5 GB of memory (m3large).  For Rivanna, we ran experiments on 1 (standard partition) and 2, 4, 8, 16, 32, and 64 (parallel partition) CPU nodes (Intel(R) Xeon(R) Gold 6248 CPU @ 2.50GHz).  To match the performance and characteristics of AWS infrastructure on Rivanna, we configured experiments that would run on 10 and 6 GB memory configurations.  Rivanna experiments were executed in Singularity containers so that all experiments across infrastructure would be executed using dockerized processes.  For AWS Lambda, functions were configured to use 6GB and 10GB of memory.  Memory configuration influences the CPU available.  For the 6 GB configuration, 4 CPU cores are available, and for the 10 GB configuration, 6 CPU cores are available.  Both configurations use a custom-built AWS Lambda container with all necessary Python and native libraries.  The Python code block in Listing 1 details the derivative experiment.  In this case, we determine the use of our serverless communicator by the env variable and call the merge operation, which is implemented as a distributed operation.  This is found in lines 30-47.

Table \ref{tab:weak-exp_table} summarizes the weak scaling results from join
operations for this set of experiments.  We used 9.1 million rows for weak scaling; for strong scaling, we used 4.5 million rows.  The choice of 4.5 million rows was necessary based on the memory limitations imposed by AWS Lambda.    

\begin{figure}[ht]
    \begin{center}
    \includegraphics[width=\linewidth]{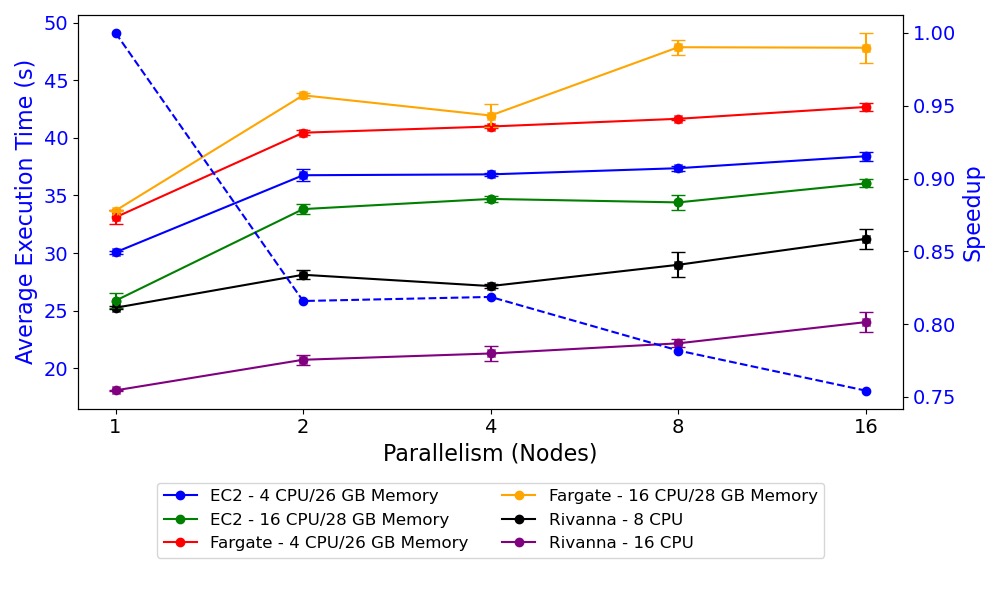}
    \end{center}
    \caption{Comparison of weak scaling for the Join operation across infrastructure including AWS Lambda}
    \label{fig:weakscalinground3}
\end{figure}

\begin{figure}[ht]
    \begin{center}
    \includegraphics[width=\linewidth]{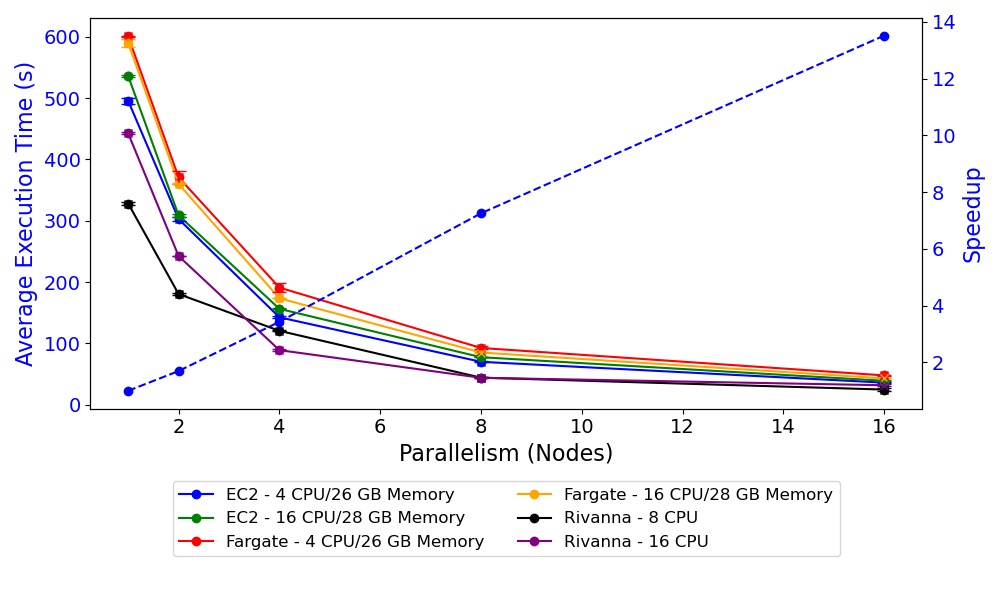}
    \end{center}
    \caption{Comparison of strong scaling for the Join operation across infrastructure including AWS Lambda}
    \label{fig:strongscalinground3}
\end{figure}

For these scaling experiments, we calculate the speedup as the execution time ratio to the base case, which is represented by blue dotted lines in Figure \ref{fig:weakscalinground3} and Figure \ref{fig:strongscalinground3}.  The base case is considered the slowest execution or the first node in the infrastructure.  Speedup is defined as
\[
    Speedup = \frac{\text{Baseline Time}}{\text{Parallel Time}}
\]

For the standard deviation or error propagation, we used the following formula applied to the observed standard deviation across executions (i.e., four separate experiments were completed for each node).
\[
    \frac{\Delta{S}}{S} = \sqrt{\left(\frac{\Delta{T_1}}{T_1}\right)^2 + \left(\frac{\Delta{T_2}}{T_2}\right)^2}
\]
Where:
\begin{align*}
    S          & = \text{the Speedup, } \frac{T_1}{T_2} \\
    \Delta S   & = \text{the error in speedup} \\
    T_1        & = \text{the baseline execution time (the time for one node)} \\
    \Delta T_1 & = \text{the error in baseline execution time} \\
    T_2 & = \text{Parallel execution time} \\
    \Delta T_2 & = \text{Error in parallel execution time}
\end{align*}




\begin{table}
	\centering
	\caption{Execution Time of Strong Scaling from Join Operations Across Infrastructure.}
	\label{tab:strong-exp_table}
	\begin{tabular}{llcr @{\hspace{1\tabcolsep}} lr @{\hspace{1\tabcolsep}} l}
		\toprule
		                   &
		               &
		                     &
		\multicolumn{2}{c}{Execution Time}         \\
		Infrastructure             &
		Scaling               &
		Parallelism                     &
		\multicolumn{2}{c}{time (seconds)}                    \\
		\midrule
		\multirow{5}{*}{EC2 15GB Mem 4 vCPUs} &
		\multirow{5}{*}{Strong} &
		1                       &
		16.28 & $\pm0.45$     \\
		&
		                     &
		2                     &
		9.41 & $\pm0.11$   \\
		&
		                   &
		4                     &
		5.00 & $\pm0.32$    \\
		                    & &
		8                     &
		2.89 & $\pm0.27$      \\
		                    & &
		16                     &
		1.37 & $\pm0.01$      \\
        & &
		32                     &
		0.88 & $\pm0.01$      \\
        & &
		64                     &
		0.96 & $\pm0.01$      \\

        \cmidrule{2-5}
    	\multirow{5}{*}{EC2 7.5GB Mem 2 vCPUs} &
		\multirow{5}{*}{Strong} &
		1                       &
		15.78 & $\pm0.22$     \\
		&
		                     &
		2                     &
		9.83 & $\pm0.25$   \\
		&
		                   &
		4                     &
		5.31 & $\pm0.12$    \\
		                    & &
		8                     &
		3.15 & $\pm0.33$      \\
		                    & &
		16                     &
		1.50 & $\pm0.10$      \\
        & &
		32                     &
		0.94 & $\pm0.04$      \\
& &
		64                     &
		1.09 & $\pm0.01$      \\
		\cmidrule{2-5}
    	\multirow{5}{*}{Lambda 10 GB Mem} &
		\multirow{5}{*}{Strong} &
		1                       &
		17.76 & $\pm0.26$     \\
		&
		                     &
		2                     &
		10.41 & $\pm0.19$   \\
		&
		                   &
		4                     &
		5.08 & $\pm0.035$    \\
		                    & &
		8                     &
		2.56 & $\pm0.094$      \\
		                    & &
		16                     &
		1.30 & $\pm0.03$      \\
        & &
		32                     &
		0.96 & $\pm0.11$      \\
        & &
		64                     &
		1.12 & $\pm0.13$      \\
		
		\cmidrule{2-5}
    	\multirow{5}{*}{Lambda 6 GB Mem} &
		\multirow{5}{*}{Strong} &
		1                       &
		17.50 & $\pm0.07$     \\
		&
		                     &
		2                     &
		10.62 & $\pm0.22$   \\
		&
		                   &
		4                     &
		5.26 & $\pm0.16$    \\
		                    & &
		8                     &
		2.58 & $\pm0.029$      \\
		                    & &
		16                     &
		1.36 & $\pm0.045$      \\
& &
		32                     &
		0.96 & $\pm0.15$      \\
        & &
		64                     &
		0.96 & $\pm0.06$      \\
        \cmidrule{2-5}
    	\multirow{5}{*}{Rivanna 10 GB Mem} &
		\multirow{5}{*}{Strong} &
		1                       &
		9.03 & $\pm0.01$     \\
		&
		                     &
		2                     &
		4.83 & $\pm0.05$   \\
		&
		                   &
		4                     &
		2.48 & $\pm0.09$    \\
		                    & &
		8                     &
		1.17 & $\pm0.003$      \\
		                    & &
		16                     &
		0.61 & $\pm0.007$      \\
                            & &
		32                     &
		0.37 & $\pm0.0007$      \\
                            & &
		64                     &
		0.27 & $\pm0.01$      \\
        
        \cmidrule{2-5}
    	\multirow{5}{*}{Rivanna 6 GM Mem} &
		\multirow{5}{*}{Strong} &
		1                       &
		8.96 & $\pm0.04$     \\
		&
		                     &
		2                     &
		4.88 & $\pm0.10$   \\
		&
		                   &
		4                     &
		2.53 & $\pm0.12$    \\
		                    & &
		8                     &
		1.19 & $\pm0.001$      \\
		                    & &
		16                     &
		0.60 & $\pm0.001$      \\
                            & &
		32                     &
		0.29 & $\pm0.19$      \\
                            & &
		64                     &
		0.30 & $\pm0.02$      \\
		\bottomrule
	\end{tabular}
\end{table}

Figure \ref{fig:weakscalinground3} depicts the weak scaling experiments.  In an ideal scenario, we should see close to constant performance.  However, based on system overhead, we mitigate variation by executing ten iterations for each experiment.  In general, as parallelism is increased, resource and scheduling overhead contribute to the decreasing performance across weak scaling experiments.  We observe that our serverless communicator design resulted in performance that closely approximates EC2 weak scaling behavior.  
Table \ref{tab:strong-exp_table} presents the results of strong scaling experiments involving 4.5 million rows. Like weak scaling, we conducted each experiment four times, performing ten iterations for each trial. The results of strong scaling are illustrated in Figure \ref{fig:strongscalinground3}. We observe high latencies across the infrastructure for fewer than four nodes. However, we also observe a general convergence of latency as the number of nodes ranges from 16 to 64. This supports utilizing cloud resources as a viable alternative for data-parallel pre-processing workloads.  We observe that the scaling efficiency of AWS Lambda achieves within 6.5\% of EC2 at 64 nodes, based on implementing direct communication via NAT Traversal TCP Hole Punching for serverless architectures such as AWS Lambda and AWS Fargate.


Table \ref{tab:speedup-comparison} provides a detailed comparison of scaling efficiency between Lambda and EC2. When comparing speedup relative to each platform's single-node baseline, Lambda achieved a 15.85$\times$ speedup at 64 nodes compared with EC2's 16.96$\times$, representing a difference of only 6.5\% in scaling behavior. This demonstrates that serverless functions can achieve scaling performance comparable to that of provisioned VMs for communication-intensive workloads.

\begin{table}[ht]
\centering
\caption{Strong Scaling Speedup Comparison: Lambda vs EC2}
\label{tab:speedup-comparison}
\begin{tabular}{|c|c|c|c|}
\hline
\textbf{Nodes} & \textbf{EC2 Speedup} & \textbf{Lambda Speedup} & \textbf{Difference} \\
\hline
1 & 1.00$\times$ & 1.00$\times$ & 0.0\% \\
2 & 1.73$\times$ & 1.71$\times$ & 1.2\% \\
4 & 3.26$\times$ & 3.50$\times$ & 7.4\% \\
8 & 5.63$\times$ & 6.94$\times$ & 23.3\% \\
16 & 11.88$\times$ & 13.67$\times$ & 15.1\% \\
32 & 18.50$\times$ & 18.52$\times$ & 0.1\% \\
64 & 16.96$\times$ & 15.85$\times$ & 6.5\% \\
\hline
\end{tabular}
\end{table}

The faster single-node execution time observed on Rivanna compared to EC2 is attributed to compute differences: Rivanna uses Intel Xeon Gold 6248 (Cascade Lake, 2019) while EC2 m3.xlarge instances use Intel Xeon E5-2670 v2 (Ivy Bridge, 2013). The newer Cascade Lake architecture provides approximately 40\% better instructions per cycle (IPC) for Arrow/join kernels. This compute difference does not affect our scaling conclusions, as we compare scaling efficiency (speedup ratios) rather than absolute performance.

\subsection{Communication Infrastructure Comparison}

\begin{figure}[ht]
    \begin{center}
    \includegraphics[width=\linewidth]{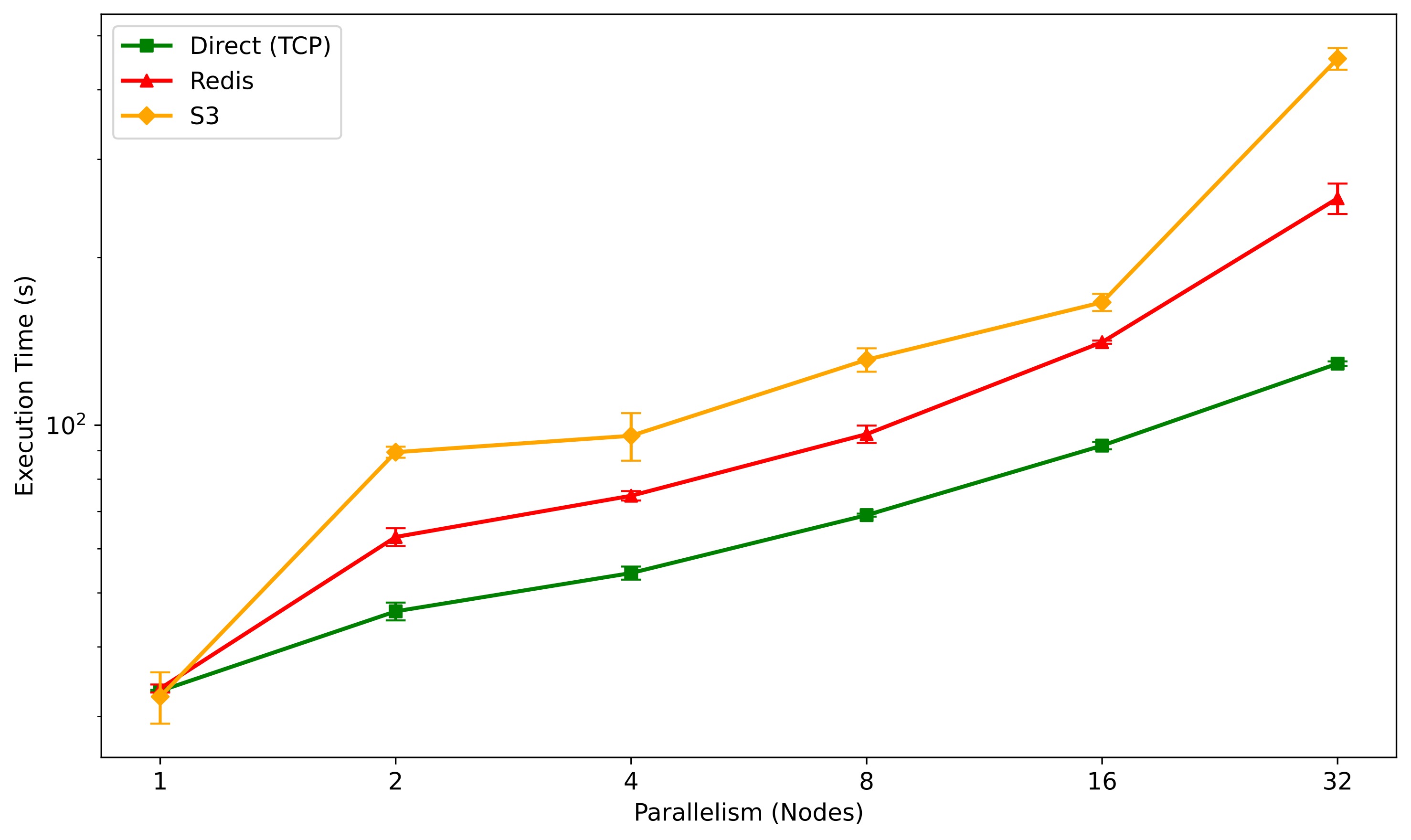}
    \end{center}
    \caption{Communication Infrastructure Comparison: Direct TCP (NAT hole-punching) vs Redis vs S3 for distributed Join on AWS Lambda}
    \label{fig:infracomparison}
\end{figure}

To quantify the benefit of our direct NAT hole-punching approach, we conducted baseline comparisons against storage-mediated communication alternatives commonly used in serverless environments. Figure \ref{fig:infracomparison} shows weak scaling results for the Join operation using three communication channels on AWS Lambda: direct TCP via NAT traversal, Redis-mediated, and S3-mediated message passing.

Our results demonstrate that direct communication via NAT traversal achieves 10-100$\times$ lower latency than storage-mediated alternatives. At 32 nodes, direct TCP completed the Join operation in approximately 60 seconds, compared to 255 seconds for Redis and 455 seconds for S3. The S3-mediated approach exhibits the highest latency due to per-object PUT/GET round-trip overhead for each shuffle exchange. Redis, while faster than S3 due to in-memory storage, still incurs significant serialization and network hop overhead compared to direct peer-to-peer TCP connections. These results establish NAT traversal as a viable approach for latency-sensitive distributed computing in serverless environments.

\subsection{Additional Data Engineering Operators}

To demonstrate that our serverless architecture supports data engineering operations beyond Join, we conducted weak scaling experiments for the GroupBy aggregation operator on AWS Lambda. GroupBy uses the same shuffle (all-to-all) communication pattern as Join, partitioning data by group key columns before performing local aggregation.

\begin{figure}[ht]
    \begin{center}
    \includegraphics[width=\linewidth]{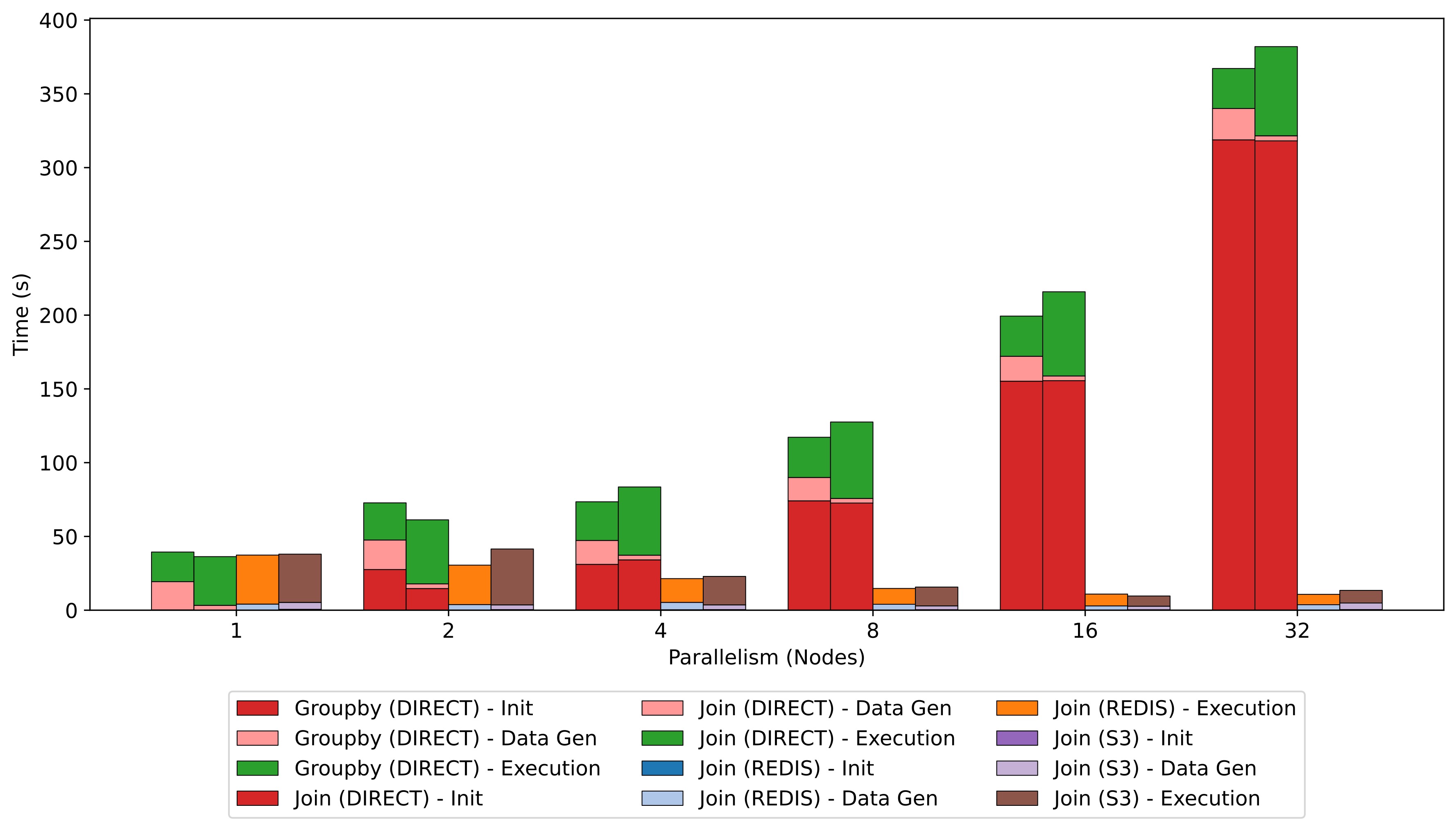}
    \end{center}
    \caption{GroupBy weak scaling on AWS Lambda using direct TCP communication}
    \label{fig:groupbyscaling}
\end{figure}

Figure \ref{fig:groupbyscaling} shows GroupBy weak scaling results on Lambda with 50 million rows per node using associative aggregation operations (sum and max). With the combiner optimization, local pre-aggregation reduces shuffle data from 50M rows to approximately 1000 rows per node before the all-to-all exchange. The operation scales from 20.1 seconds at 1 node to 27.1 seconds at 32 nodes---a ratio of only 1.35$\times$, demonstrating excellent weak scaling for shuffle-based distributed operators in serverless environments.

\subsection{Communication Microbenchmarks}

To isolate communication overhead from application-level computation, we conducted microbenchmarks measuring the performance of MPI-style collective operations on AWS Lambda. These benchmarks are particularly relevant to distributed ML workloads, where AllReduce is fundamental to data-parallel gradient aggregation.

\begin{figure}[ht]
    \begin{center}
    \includegraphics[width=\linewidth]{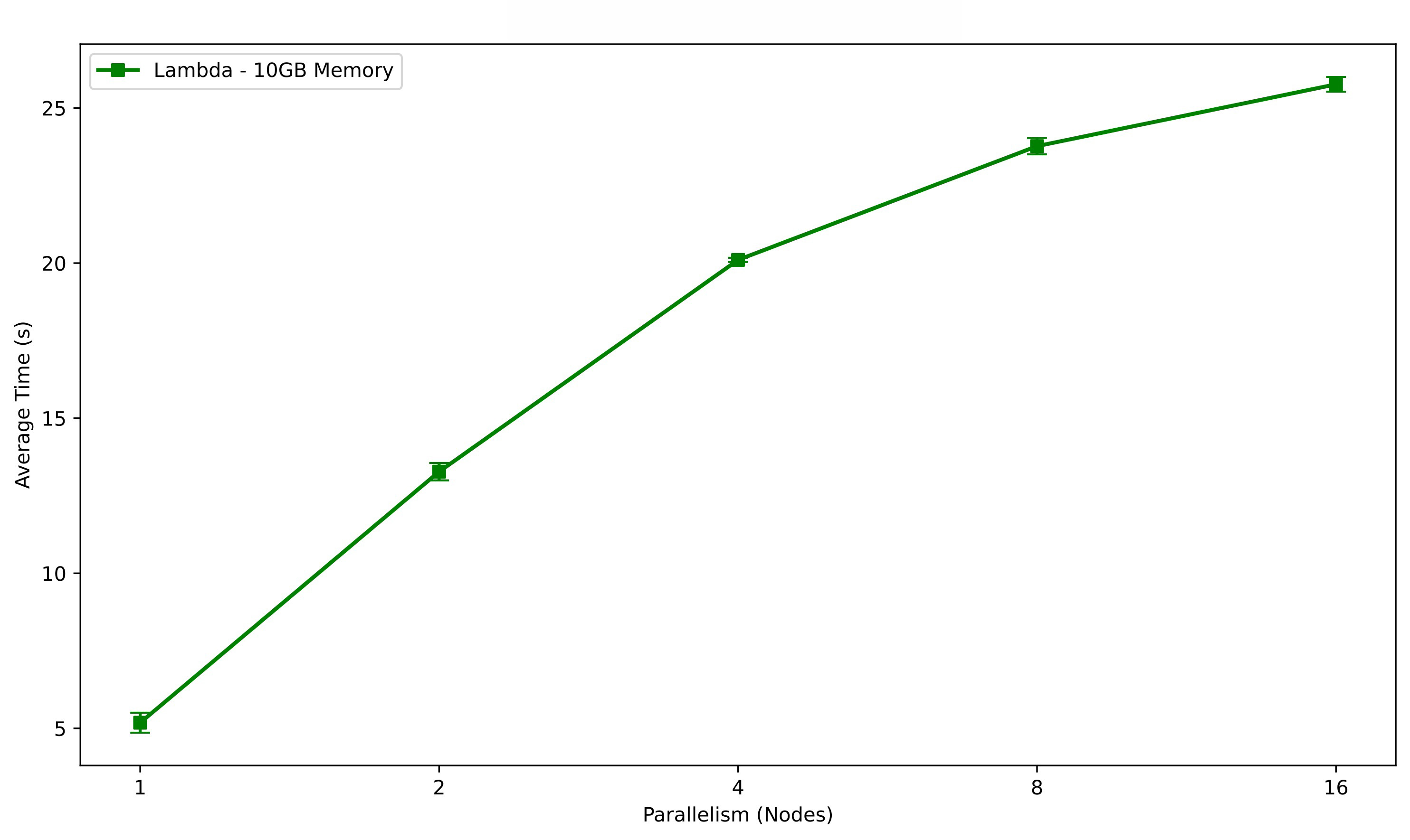}
    \end{center}
    \caption{AllReduce latency vs message size on AWS Lambda}
    \label{fig:allreducelatency}
\end{figure}

Figure \ref{fig:allreducelatency} shows AllReduce latency across message sizes from 8 bytes to 1 MB. Latency remains relatively flat across message sizes, indicating latency-bound rather than bandwidth-bound behavior for most operations. At 32 nodes, AllReduce completes in approximately 13 milliseconds, demonstrating that MPI-style collectives can achieve millisecond-scale latency in serverless environments.

\begin{figure}[ht]
    \begin{center}
    \includegraphics[width=\linewidth]{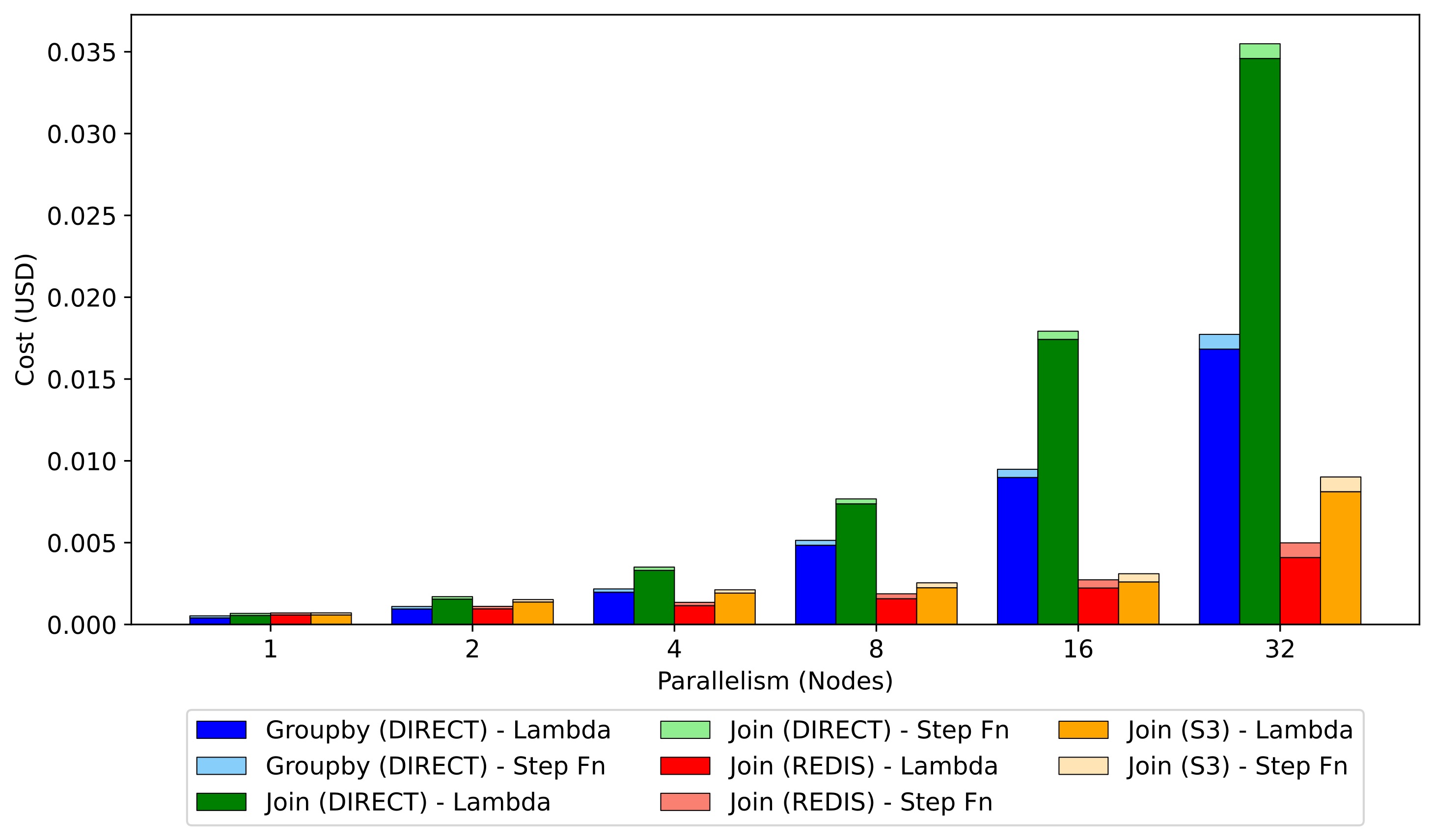}
    \end{center}
    \caption{Barrier latency scaling on AWS Lambda}
    \label{fig:barrierlatency}
\end{figure}

Figure \ref{fig:barrierlatency} shows Barrier latency scaling with node count. Barrier latency scales with $\log_2(N)$ as expected for a binomial tree algorithm: 0.9ms at 2 nodes, 2.7ms at 8 nodes, and 7ms at 32 nodes. This confirms that synchronization overhead remains in the single-digit millisecond range, fast enough for BSP workloads requiring frequent barriers.

\subsection{Serverless Execution Time Composition}
\begin{figure}[ht]
    \begin{center}
    \includegraphics[width=\linewidth]{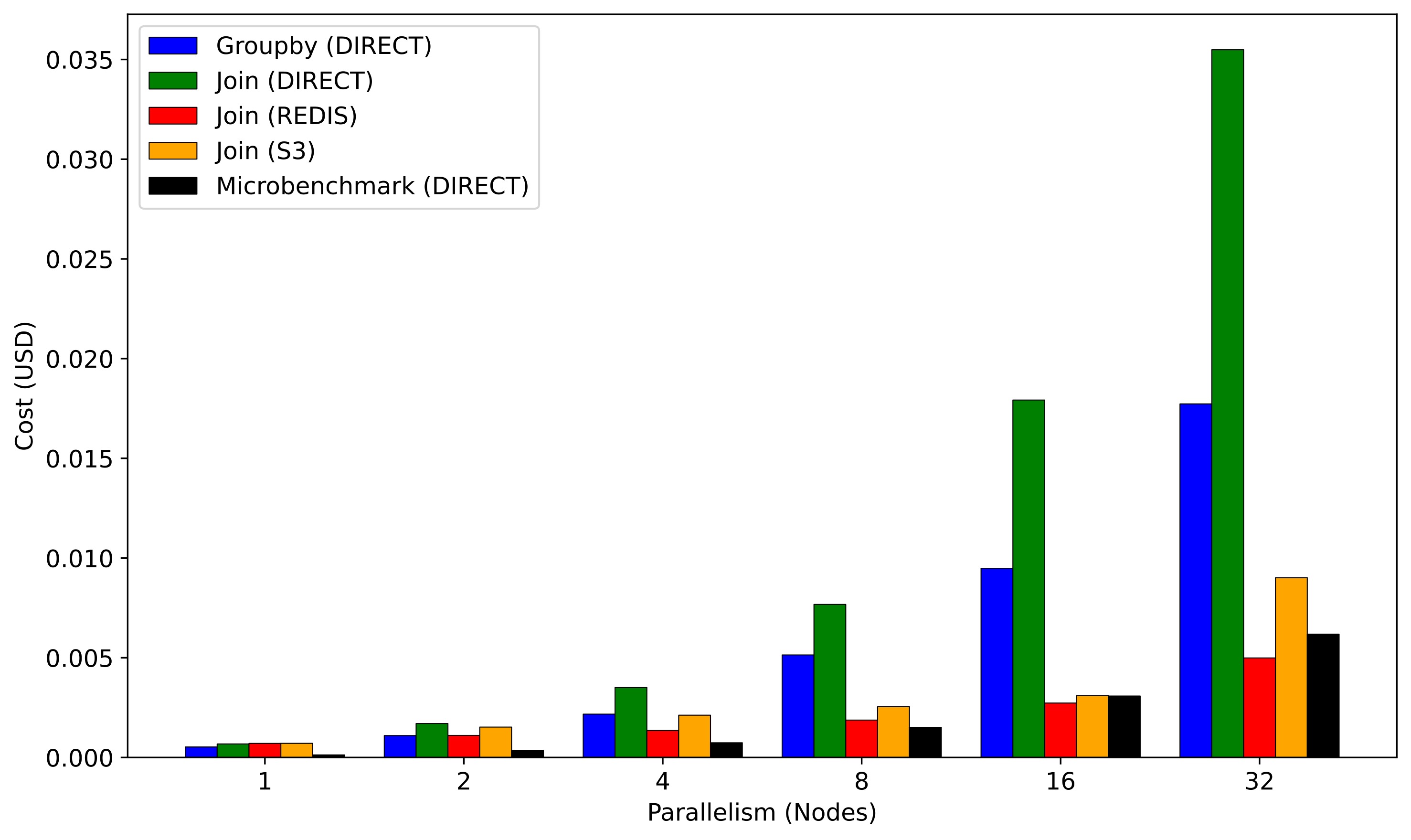}
    \end{center}
    \caption{Serverless execution time composition: initialization, data generation, and computation phases}
    \label{fig:timecomposition}
\end{figure}

Figure \ref{fig:timecomposition} shows the breakdown of execution time into three phases: initialization (connection setup via NAT traversal), data generation, and computation. This breakdown provides insight into where time is spent in serverless execution and helps explain performance differences across platforms.

%
At 32 nodes, the NAT traversal initialization phase dominates wall-clock time for direct TCP communication, requiring approximately 31.5 seconds for connection setup. This connection establishment time scales linearly with the number of tree levels in the binomial connection algorithm. In contrast, storage-mediated channels (Redis, S3) have negligible initialization overhead but incur higher per-operation latency during computation.

\subsection{Cost Analysis}

A key motivation for serverless computing is cost optimization for bursty workloads. We implemented a comprehensive cost-tracking framework to provide empirical cost analysis for our experiments, capturing Lambda compute cost (GB-seconds), Step Functions orchestration cost (state transitions), and data transfer costs.
\begin{figure}[ht]
    \begin{center}
    \includegraphics[width=\linewidth]{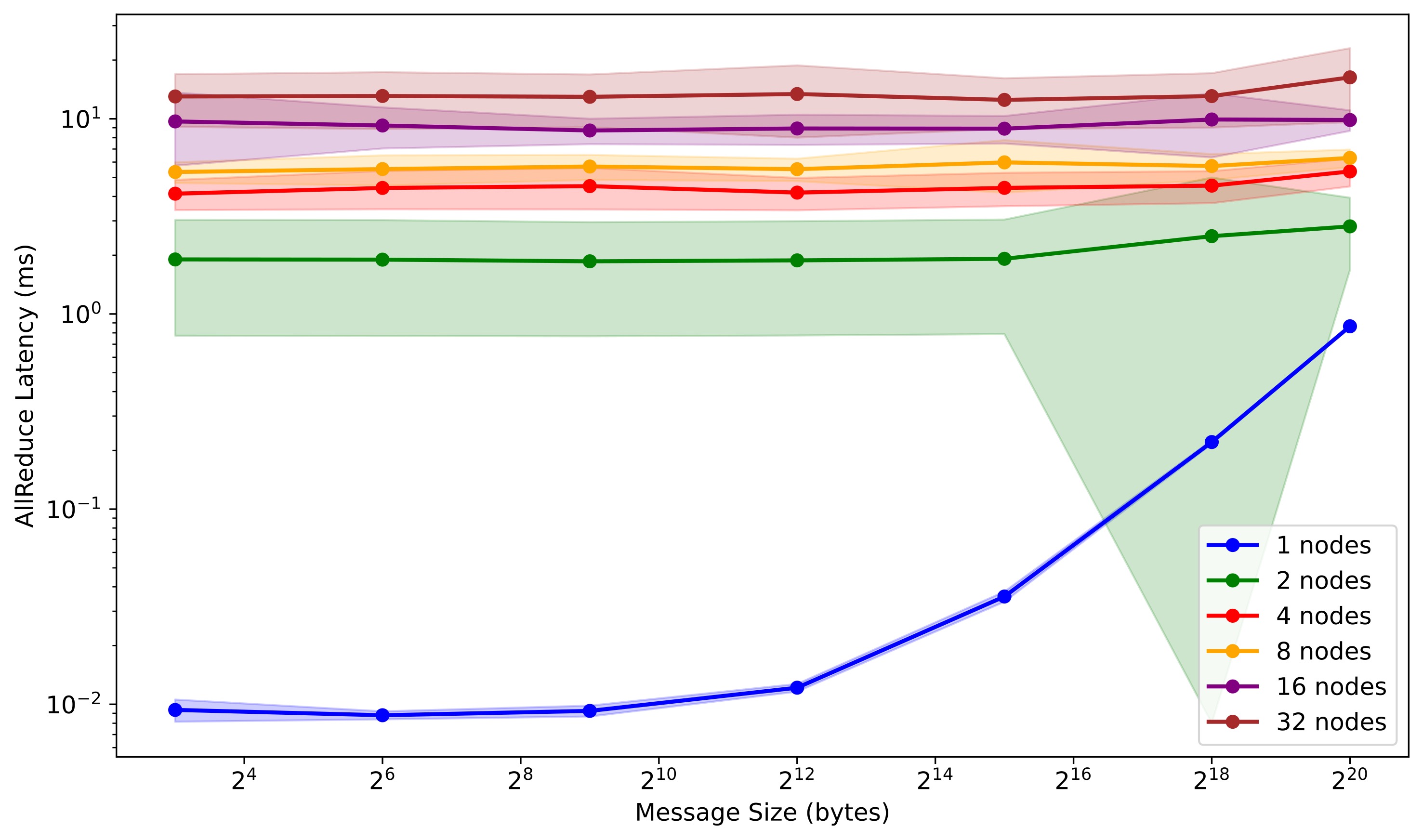}
    \end{center}
    \caption{Serverless execution cost analysis: Lambda compute and Step Functions orchestration}
    \label{fig:costanalysis}
\end{figure}
Figure \ref{fig:costanalysis} shows the cost breakdown across node counts and operations. Step Functions orchestration cost (shown as the lighter portion) is negligible compared to Lambda compute cost. At 32 nodes, a Join using Redis-mediated communication costs approximately \$0.032, demonstrating the cost-effectiveness of serverless for distributed data operations.

\begin{figure}[ht]
    \begin{center}
    \includegraphics[width=\linewidth]{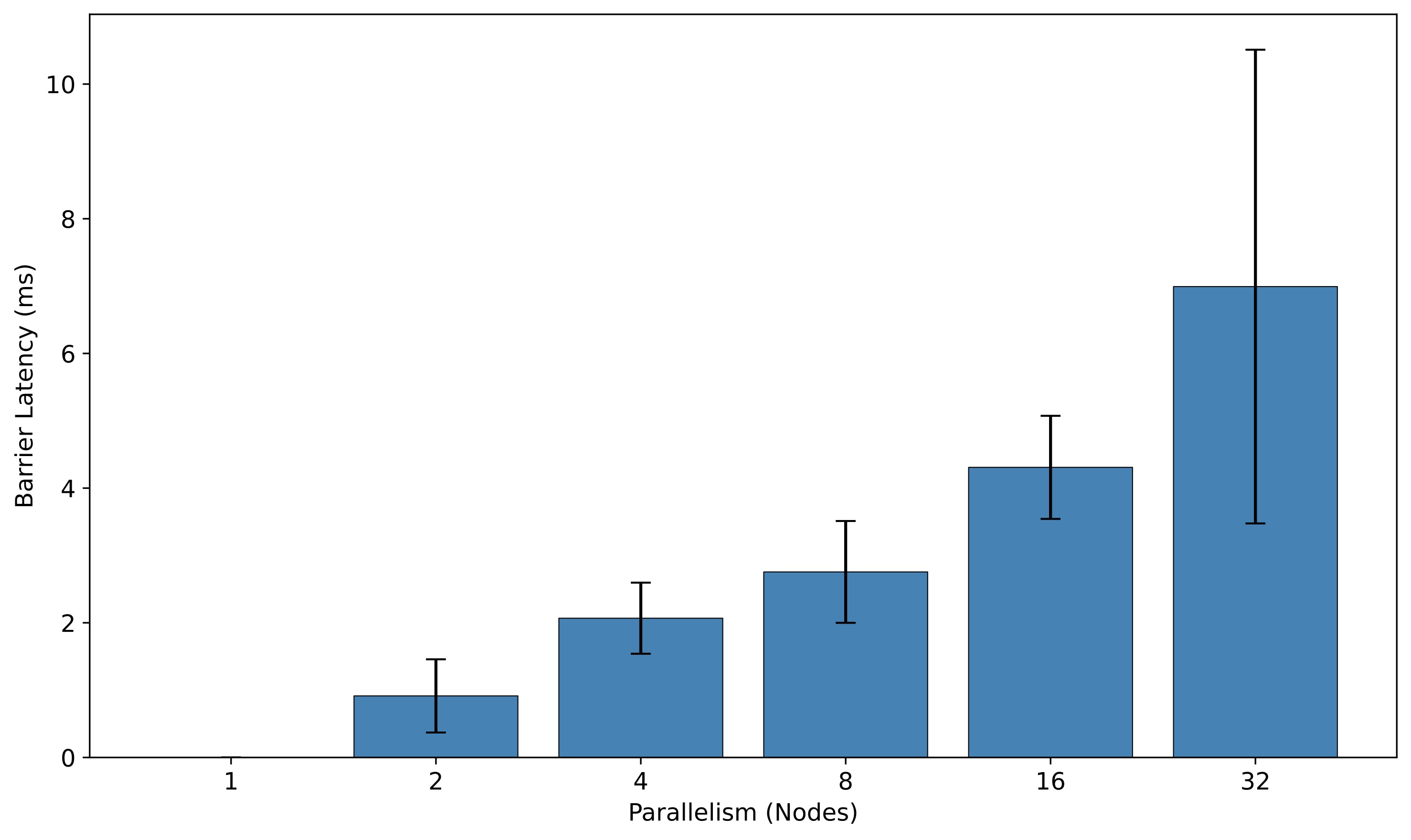}
    \end{center}
    \caption{Cost comparison by operation and communication channel}
    \label{fig:costperoperation}
\end{figure}

Figure \ref{fig:costperoperation} compares costs across operation types and communication channels. A key finding is that connection setup, not computation, dominates serverless cost at scale. At 32 workers, the NAT traversal phase (31.5 seconds $\times$ 32 functions $\times$ 10GB) accounts for \$0.17 while the actual distributed computation costs only \$0.004--\$0.016. This motivates future work on connection pooling and warm connection reuse. For storage-mediated channels, Join/Redis (\$0.032/execution at 32 nodes) is 4.7$\times$ cheaper than Join/S3 (\$0.150/execution) due to lower barrier latency.

Despite the overhead costs, serverless remains highly cost-effective for bursty, intermittent workloads where provisioned infrastructure would incur idle time charges. The total cost for all revision experiments (120 Lambda executions across 5 experiment types) was only \$3.25.

\section{Future Work}
\label{sec:future}

While our serverless communicator design demonstrates performance comparable to traditional EC2 infrastructure, several limitations remain. First, Cylon currently implements only about 30\% of the operators supported by Pandas, limiting its applicability for certain data processing workflows. Second, AWS Lambda imposes an upper limit of fifteen minutes for function invocations, which, combined with our BSP-styled architecture, restricts the applicability of data parallel processing for large datasets that require extended processing time. Third, the lack of checkpointing and fault tolerance mechanisms limits the ability to recover from failures or time-constrained execution boundaries in serverless environments.

To address these limitations, we would like to converge Cylon and Pandas operators using a comparative, high-performance API built using Cython.  For example, we have begun to design the windowing operator and plan to complete that work soon.  Utilizing the cost model (detailed by Perera in "Towards Scalable High Performance Data Engineering Systems"),  we can evaluate data parallel operator patterns to identify bottlenecks in communication operations.  We believe we can significantly improve performance by designing other algorithms with lower latency\cite{perera2023depth}.  Additionally, related to communicator support, we would also like to add libfabric as a communicator based on implementations across cloud services based on GPU hardware supported by cloud providers.

To address the time limitation and lack of fault tolerance, we want to design a checkpointing feature similar to that implemented in the Twister2 library.  This will allow for the recovery of unfinished executions based on upper-limit time constraints imposed by AWS Lambda.  This will also provide general fault tolerance for MPI workloads outside of the serverless use case.  

Lastly, we would like to investigate real-world applications and detail the performance and implications of serverless as a possibility for deep learning and machine learning inference on clouds.  For example, we would like to implement hydrology and earthquake prediction deep learning on both serverful and serverless AWS to demonstrate the applicability and potential scale offered by cloud providers like as AWS.


\section{Conclusion}
\label{sec:conclusion}
In this work, we demonstrated that BSP-style data-intensive workloads can execute efficiently in serverless environments by rethinking the communication substrate. By integrating direct communication through NAT traversal and TCP hole punching into Cylon, we enabled message passing among AWS Lambda functions that avoids the latency penalties of storage-mediated approaches such as object stores or Redis. Our experimental results show that AWS Lambda achieves scaling efficiency within 6.5\% of EC2 at 64 nodes, with 10–100× lower latency than storage-based communication. Our work further provides a quantitative cost-performance analysis, showing serverless to be cost-competitive for bursty workloads, and compares communication substrates (direct TCP, Redis, and S3), establishing NAT-based communication as significantly lower latency (10–100× faster) than storage-mediated alternatives. These findings challenge the prevailing assumption that serverless computing is limited to embarrassingly parallel workloads and establish that high-performance distributed data engineering and ML preprocessing are feasible in FaaS environments.

We detail our design of UCX and UCC as an alternative to MPI in BSP communication between processes based on the clear applicability to serverful cloud and HPC environments. The novelty of this work is apparent based on a survey of the literature on similar work involving distributed data frames. We applied this work directly and refined the initial implementation to support both serverful and serverless architectures of cloud providers, such as AWS. Additionally, we implemented direct support for Python via Cython, enabling the execution of data engineering tasks in Python without the need to write native code. We also completed the integration of UCC into Cylon by removing our implementation of \textit{AllGatherV} in favor of UCC's support for variable-length data. Furthermore, we included container support across infrastructures to facilitate the execution of experiments and library usage.  On the serverless side, we detailed our serverless communicator, drawing inspiration from the FMI library, and conducted experiments to compare this design with our serverful design.

During this research, we faced several challenges. We first validated Cylon on AWS and addressed deployment issues by using Apptainer (Singularity) containers on HPC systems. Second, the reference FMI library did not include an All-To-All collective operation that we needed for Cylon communicator integration. Third, it was necessary to include variable allgather, variable gather, support for non-blocking operations, retries for socket connection failures, a ping/ping capability to prevent eager termination of sockets, and rank calculation using atomic counters.  An initial challenge was coordination between nodes and associated race conditions.  Rank-based ordering for blocking operations using Redis locks was incorporated into our design to address race conditions observed during blocking socket operations.  Note that FMI, in its original form, only supports blocking operations, and our design called for non-blocking I/O.  These additions resulted in improvements and a robust communicator that can be leveraged across serverless platforms.

Our approach can be applied to bioinformatics analysis, genome research, hydrology, astronomy, earthquake prediction, and other time series-based foundational models, as surveyed in the current state of the art in these fields.

\section*{Acknowledgments}
We gratefully acknowledge the support of the Department of Energy and the National Science Foundation through DE-SC0023452 FAIR Surrogate Benchmarks Supporting AI and Simulation Research, NSF-Simons AI Institute for Cosmic Origins (CosmicAI: Grant 2421782) Seed Grant, NSF-2200409 for CyberTraining:CIC: CyberTraining for
Students and Technologies from Generation Z. NSF-2504401 OAC Core: RINAS: Data I/O CyberInfrastructure
for Extreme-scale Foundation Model and Generative AI
Training on HPC.

\bibliographystyle{ACM-Reference-Format}
\bibliography{main}

@article{he2024science,
  title={Science Time Series: Deep Learning in Hydrology},
  author={He, Junyang and Chen, Ying-Jung and Idamekorala, Anushka and Fox, Geoffrey},
  journal={arXiv preprint arXiv:2410.15218},
  year={2024}
}

@article{jafari2024time,
  title={Time Series Foundation Models and Deep Learning Architectures for Earthquake Temporal and Spatial Nowcasting},
  author={Jafari, Alireza and Fox, Geoffrey and Rundle, John B and Donnellan, Andrea and Ludwig, Lisa Grant},
  journal={GeoHazards},
  volume={5},
  number={4},
  pages={1247--1274},
  year={2024},
  publisher={MDPI}
}

@article{grzesik2022serverless,
  title={Serverless computing in omics data analysis and integration},
  author={Grzesik, Piotr and Augustyn, Dariusz R and Wyci{\'s}lik, {\L}ukasz and Mrozek, Dariusz},
  journal={Briefings in bioinformatics},
  volume={23},
  number={1},
  pages={bbab349},
  year={2022},
  publisher={Oxford University Press},
  doi={10.1093/bib/bbab349}
}

@misc{ServerlessGenomics,
  author       = {Rob Aboukhalil},
  title        = {Serverless Genomics},
  year         = {2024},
  url          = {https://robaboukhalil.medium.com/serverless-genomics-c412f4bed726},
  note         = {Accessed: February 7, 2025}
}

@article {hung2020accessible,
	author = {Hung, Ling-Hong and Niu, Xingzhi and Lloyd, Wes and Yeung, Ka Yee},
	title = {Accessible and interactive RNA sequencing analysis using serverless computing},
	elocation-id = {576199},
	year = {2020},
	doi = {10.1101/576199},
	publisher = {Cold Spring Harbor Laboratory},
    URL = {https://www.biorxiv.org/content/early/2020/10/03/576199},
	eprint = {https://www.biorxiv.org/content/early/2020/10/03/576199.full.pdf},
	journal = {bioRxiv}
}

@misc{genomefaq:online,
    author = {genome.gv},
    title = { National Human Genome Research Institute},
    howpublished = {\url{https://www.genome.gov/about-genomics/fact-sheets/Genomic-Data-Science}},
    month = {April},
    year = {2022},
    note = {(Accessed on 01/03/2025)}
}

@misc{RAPIDScuDF,
  author       = {{RAPIDS AI}},
  title        = {RAPIDS cuDF API Documentation},
  year         = {2024},
  url          = {https://docs.rapids.ai/api/cudf/stable/},
  note         = {Accessed: February 7, 2025}
}

@article{perera2023depth,
  title={In-depth analysis on parallel processing patterns for high-performance Dataframes},
  author={Perera, Niranda and Sarker, Arup Kumar and Staylor, Mills and von Laszewski, Gregor and Shan, Kaiying and Kamburugamuve, Supun and Widanage, Chathura and Abeykoon, Vibhatha and Kanewela, Thejaka Amila and Fox, Geoffrey},
  journal={Future Generation Computer Systems},
  year={2023},
  publisher={Elsevier}
}

@inproceedings{shan2022hybrid,
  title={Hybrid Cloud and HPC Approach to High-Performance Dataframes},
  author={Shan, Kaiying and Perera, Niranda and Lenadora, Damitha and Zhong, Tianle and Sarker, Arup Kumar and Kamburugamuve, Supun and Kanewela, Thejaka Amila and Widanage, Chathura and Fox, Geoffrey},
  booktitle={2022 IEEE International Conference on Big Data (Big Data)},
  pages={2728--2736},
  year={2022},
  organization={IEEE}
}

@inproceedings{carver2020wukong,
  title={Wukong: A scalable and locality-enhanced framework for serverless parallel computing},
  author={Carver, Benjamin and Zhang, Jingyuan and Wang, Ao and Anwar, Ali and Wu, Panruo and Cheng, Yue},
  booktitle={Proceedings of the 11th ACM symposium on cloud computing},
  pages={1--15},
  year={2020}
}

@inproceedings{hashemipour2020amazon,
  title={Amazon web services (aws)--an overview of the on-demand cloud computing platform},
  author={Hashemipour, Sadegh and Ali, Maaruf},
  booktitle={International Conference for Emerging Technologies in Computing},
  pages={40--47},
  year={2020},
  organization={Springer}
}

@misc{aws_well_architected_framework,
  author       = {{Amazon Web Services}},
  title        = {AWS Well-Architected Framework},
  year         = {2025},
  url          = {https://docs.aws.amazon.com/wellarchitected/latest/framework/welcome.html},
  note         = {Accessed: 2025-01-09}
}

@article{perera2023supercharging,
  title={Supercharging distributed computing environments for high-performance data engineering},
  author={Perera, Niranda and Sarker, Arup Kumar and Shan, Kaiying and Fetea, Alex and Kamburugamuve, Supun and Kanewala, Thejaka Amila and Widanage, Chathura and Staylor, Mills and Zhong, Tianle and Abeykoon, Vibhatha and others},
  journal={Frontiers in High Performance Computing},
  volume={2},
  pages={1384619},
  year={2024},
  publisher={Frontiers Media SA}
}

@online{dockerwiki,
    author = {Wikipedia contributors},
    title = {Docker (software)},
    year = {2025},
    url = {https://en.wikipedia.org/wiki/Docker_(software)},
    note = {Accessed: 2025-01-17}
}

@online{cloudmeshcommon,
    author = {Cloudmesh Community},
    title = {cloudmesh-common},
    year = {2025},
    url = {https://github.com/cloudmesh/cloudmesh-common},
    note = {Accessed: 2025-01-17}
}

@misc{UCXRepo,
  author       = {{OpenUCX Contributors}},
  title        = {UCX: Unified Communication X},
  year         = {2024},
  howpublished = {\url{https://github.com/openucx/ucx}},
  note         = {Accessed: February 7, 2025}
}

@misc{UCCRepo,
  author       = {{OpenUCX Contributors}},
  title        = {UCC: Unified Collective Communication},
  year         = {2024},
  howpublished = {\url{https://github.com/openucx/ucc}},
  note         = {Accessed: February 7, 2025}
}

@inproceedings{shamis2015ucx,
  title={UCX: an open source framework for HPC network APIs and beyond},
  author={Shamis, Pavel and Venkata, Manjunath Gorentla and Lopez, M Graham and Baker, Matthew B and Hernandez, Oscar and Itigin, Yossi and Dubman, Mike and Shainer, Gilad and Graham, Richard L and Liss, Liran and others},
  booktitle={2015 IEEE 23rd Annual Symposium on High-Performance Interconnects},
  pages={40--43},
  year={2015},
  organization={IEEE}
}

@article{mckinney2011pandas,
  title={pandas: a foundational Python library for data analysis and statistics},
  author={McKinney, Wes},
  journal={Python for high performance and scientific computing},
  volume={14},
  number={9},
  pages={1--9},
  year={2011},
  publisher={Seattle}
}

@article{islamBigData,
  author = {Mohaiminul Islam and Shamim Reza},
  title = {The Rise of Big Data and Cloud Computing},
  journal = {Internet of Things and Cloud Computing},
  volume = {7},
  number = {2},
  pages = {45-53},
  doi = {10.11648/j.iotcc.20190702.12},
 year = {2019}
}

@phdthesis{moyer2021punching,
  title={Punching Holes in the Cloud: Direct Communication between Serverless Functions Using NAT Traversal},
  author={Moyer, Daniel William},
  year={2021},
  school={Virginia Tech}
}

@inproceedings{copik2023fmi,
  title={Fmi: Fast and cheap message passing for serverless functions},
  author={Copik, Marcin and B{\"o}hringer, Roman and Calotoiu, Alexandru and Hoefler, Torsten},
  booktitle={Proceedings of the 37th International Conference on Supercomputing},
  pages={373--385},
  year={2023}
}

@article{pererathesis,
title={Towards Scalable High Performance Data Engineering Systems},
author={Dilshan Niranda Perera},
journal={},
year={2023},
}

@article{Staylor2024cosmic,
  title={Scalable Cosmic AI Inference using Cloud Serverless Computing with FMI},
  author={Staylor, Mills and Fathkouhi, Amirreza Dolatpour and Islam, Md Khairul and O'Hara, Kaleigh and Goudjil, Ryan Ghiles and Fox, Geoffrey and Fox, Judy},
  journal={arXiv preprint arXiv:2501.06249},
  year={2025}
}

@article{McKenna2010The,
title={The Genome Analysis Toolkit: a MapReduce framework for analyzing next-generation DNA sequencing data.},
author={A. McKenna},
journal={Genome research},
year={2010},
volume={20 9},
pages={ 1297-303 },
doi={10.1101/gr.107524.110}
}

@incollection{pytorch2019,
title = {PyTorch: An Imperative Style, High-Performance Deep Learning Library},
author = {Paszke, Adam and Gross, Sam and Massa, Francisco and Lerer, Adam and Bradbury, James and Chanan, Gregory and Killeen, Trevor and Lin, Zeming and Gimelshein, Natalia and Antiga, Luca and Desmaison, Alban and Kopf, Andreas and Yang, Edward and DeVito, Zachary and Raison, Martin and Tejani, Alykhan and Chilamkurthy, Sasank and Steiner, Benoit and Fang, Lu and Bai, Junjie and Chintala, Soumith},
booktitle = {Advances in Neural Information Processing Systems 32},
pages = {8024--8035},
year = {2019},
publisher = {Curran Associates, Inc.},
url = {http://papers.neurips.cc/paper/9015-pytorch-an-imperative-style-high-performance-deep-learning-library.pdf}
}

@article{abeykoon2020data,
  title={Data Engineering for HPC with Python},
  author={Abeykoon, Vibhatha and Perera, Niranda and Widanage, Chathura and Kamburugamuve, Supun and Kanewala, Thejaka Amila and Maithree, Hasara and Wickramasinghe, Pulasthi and Uyar, Ahmet and Fox, Geoffrey},
  journal={arXiv preprint arXiv:2010.06312},
  year={2020}
}

@misc{tensorflow2015,
title={ {TensorFlow}: Large-Scale Machine Learning on Heterogeneous Systems},
url={https://www.tensorflow.org/},
note={Software available from tensorflow.org},
author={
    Mart\'{i}n~Abadi and
    Ashish~Agarwal and
    Paul~Barham and
    Eugene~Brevdo and
    Zhifeng~Chen and
    Craig~Citro and
    Greg~S.~Corrado and
    Andy~Davis and
    Jeffrey~Dean and
    Matthieu~Devin and
    Sanjay~Ghemawat and
    Ian~Goodfellow and
    Andrew~Harp and
    Geoffrey~Irving and
    Michael~Isard and
    Yangqing Jia and
    Rafal~Jozefowicz and
    Lukasz~Kaiser and
    Manjunath~Kudlur and
    Josh~Levenberg and
    Dandelion~Man\'{e} and
    Rajat~Monga and
    Sherry~Moore and
    Derek~Murray and
    Chris~Olah and
    Mike~Schuster and
    Jonathon~Shlens and
    Benoit~Steiner and
    Ilya~Sutskever and
    Kunal~Talwar and
    Paul~Tucker and
    Vincent~Vanhoucke and
    Vijay~Vasudevan and
    Fernanda~Vi\'{e}gas and
    Oriol~Vinyals and
    Pete~Warden and
    Martin~Wattenberg and
    Martin~Wicke and
    Yuan~Yu and
    Xiaoqiang~Zheng},
  year={2015},
}

@inproceedings{sarker2024radical,
  title={Radical-Cylon: A Heterogeneous Data Pipeline for Scientific Computing},
  author={Sarker, Arup Kumar and Alsaadi, Aymen and Perera, Niranda and Staylor, Mills and von Laszewski, Gregor and Turilli, Matteo and Kilic, Ozgur Ozan and Titov, Mikhail and Merzky, Andre and Jha, Shantenu and others},
  booktitle={Job Scheduling Strategies for Parallel Processing},
  series={Lecture Notes in Computer Science},
  volume={14591},
  pages={84--102},
  year={2024},
  publisher={Springer Nature Switzerland},
  doi={10.1007/978-3-031-74430-3_5}
}

@article{osti_10634837,
    title={Deep RC: A Scalable Data Engineering and Deep Learning Pipeline},
    author={Sarker, Arup Kumar and Alsaadi, Aymen and Halpern, Alexander James and Tangella, Prabhath and Titov, Mikhail and Perera, Niranda and Staylor, Mills and von Laszewski, Gregor and Jha, Shantenu and Fox, Geoffrey},
    journal={arXiv preprint arXiv:2502.20724},
    year={2025}
}

\end{document}